\documentclass[aps,prd,preprint,tightenlines,nofootinbib]{revtex4}

\pdfoutput=1

\usepackage{amsfonts,amsmath,amsthm,amssymb}
\usepackage{mathrsfs}
\usepackage{bm}
\usepackage{color}
\usepackage{epsfig}

\usepackage[usenames,dvipsnames]{xcolor}
\definecolor{orange}{cmyk}{0,0.5,1,0}
\definecolor{rossoCP3}{cmyk}{0,.88,.77,.40}
\definecolor{graa}{rgb}{0.8,0.8,0.8}
\definecolor{blaa}{rgb}{0.2,0.2,0.6}

\newcommand{\PRE}[1]{{#1}}   

\newcommand{\beq}{\begin{equation}}
\newcommand{\eeq}{\end{equation}}
\newcommand{\bea}{\begin{flushleft} \begin{eqnarray}}
\newcommand{\eea}{\end{eqnarray}\end{flushleft}}

\newcommand{\comment}[1]{}

\newcommand{\ci}[1]{}

\newcommand{\pd}{\partial}

\newcommand{\ba}{\begin{eqnarray}}
\newcommand{\ea}{\end{eqnarray}}
\newcommand{\be}{\begin{equation}}
\newcommand{\ee}{\end{equation}}
\newcommand{\bay}[1]{\left(\begin{array}{#1}}
\newcommand{\eay}{\end{array}\right)}

\def\CM{{\cal M}}

\def\CV{{\cal V}}

\newcommand{\beqa}{\begin{eqnarray}}
\newcommand{\eeqa}{\end{eqnarray}}

\newcommand{\la}{\langle}
\newcommand{\ra}{\rangle}
\newcommand{\lpa}{\left(}
\newcommand{\rpa}{\right)}

\begin{document}

\title{\PRE{\vspace*{0.9in}} \color{rossoCP3}{
$\bm{W}$-WIMP  Annihilation as a Source of the  \emph{Fermi} Bubbles{}}
\PRE{\vspace*{0.1in}} }

\author{Luis Alfredo Anchordoqui}
\email{luis.anchordoqui@gmail.com}
\affiliation{Department of Physics,\\ 
University of Wisconsin-Milwaukee,  Milwaukee, WI 53201, USA
\PRE{\vspace*{.05in}}
}

\author{Brian James Vlcek}
\email{brian.vlcek@gmail.com}
\affiliation{Department of Physics,\\
University of Wisconsin-Milwaukee,  Milwaukee, WI 53201, USA
\PRE{\vspace*{.05in}}
}

\date{May 2013}

\PRE{\vspace*{.15in}}

\begin{abstract}\vskip 2mm
  \noindent The Fermi Gamma-ray Space Telescope discovered two
  $\gamma$-ray emitting bubble-shaped structures that extend nearly
  symmetrically on either side of our Galaxy and appear
  morphologically connected to the Galactic center. The origin of the
  emission is still not entirely clear. It was recently shown that
  the spectral shape of the emission from the \emph{Fermi} bubbles is
  well described by an approximately $50~{\rm GeV}$ dark matter
  particle annihilating to $b \bar b$, with a normalization
  corresponding to a velocity average annihilation cross section $\langle
  \sigma_b \, v \rangle \approx 8 \times 10^{-27} {\rm cm}^3/{\rm s}$.
  We study the minimal hidden sector recently introduced by Weinberg
  and examine to what extent its weakly interacting massive particles
  ($W$-WIMPs) are capable of accommodating both the desired effective
  annihilation rate into quarks and the observed dark matter density.
\end{abstract}

\maketitle

\section{Introduction}

Beyond standard model (SM) physics models to be probed at the Large
Hadron Collider (LHC) often include the concept of a hidden sector,
consisting of $SU(3) \times SU(2) \times U(1)$ singlet
fields. Independent of any model, the standard sector and the hidden
sector are coupled by interactions of gauge-invariant operators which
illuminate the path for exploring structures in the hidden sector by
observing phenomena in the visible standard sector. A tantalizing
realization of this idea is provided by the Higgs portal, which
connects the Higgs fields in the two sectors by an elementary quartic
interaction~\cite{Schabinger:2005ei,Patt:2006fw,Barger:2007im,Barger:2008jx,Andreas:2010dz,Logan:2010nw,Bock:2010nz,Englert:2011yb}. Such
a construct moves a precision study of the Higgs sector into a central
position of new physics searches at the LHC. Likewise, astrophysical
observations open the gates for complementary information to further
test the Higgs portal hypothesis and to improve our understanding of
the physics in the hidden sector.

Around the Galactic center (GC), there exists a bright and spatially
extended source \mbox{$\gamma$-ray} emission peaking at energies of a few
GeV. The spectrum and morphology of this signal is consistent with one 
originating from dark matter
annihilations~\cite{Goodenough:2009gk,Hooper:2010mq,Hooper:2011ti,Abazajian:2012pn}. Very
recently, evidence of this signal has been found from regions outside
of the GC~\cite{Hooper:2013rwa} in the directions of the sky
coincident with the \emph{Fermi} bubbles: two bilateral $\gamma$-ray
lobes centered at the core of the Galaxy and extending to around
$50^\circ$ above and below the Galactic plane ({\it i.e.}, $r =
\pm 10~{\rm kpc}$, where $r$ is the distance from the
GC)~\cite{Dobler:2009xz,Su:2010qj}. At lower Galactic latitudes, these
structures are coincident with a nonthermal microwave ``haze'' found in
WMAP $23-33$~GHz data~\cite{Finkbeiner:2003im} (confirmed recently by
the Planck space mission~\cite{:2012rta}) and the thermal x-ray
emission seen by ROSAT~\cite{Snowden:1997ze}.

Far from the Galactic plane ($|b|\agt 30^{\circ}$), the observed
energy-weighted $\gamma$-ray spectrum is nearly invariant with
latitude and fairly flat ($d\Phi_\gamma/dE_\gamma \propto E_\gamma^{-2}$) over
the energy range observed by \emph{Fermi}. The correlation found in
the multiwavelength observations seems to indicate that the bubbles
(measured in the range of $E_\gamma \sim 1 - 100~{\rm GeV}$) are
produced by a population of ${\rm GeV}-{\rm TeV}$ electrons (with an
approximately power-law spectrum $d\Phi_e/dE_e \propto E_e^{-3}$) via
inverse Compton scattering of ambient low-energy photons, as the same
electrons can also simultaneously produce radio synchrotron radiations
in the presence of magnetic
fields~\cite{Dobler:2009xz,Mertsch:2011es}. The transparency of
  this elementary and self-consistent framework provides strong
  support for a leptonic origin of the high-latitude emission from the
  \emph{Fermi} bubbles.

Conversely, at latitudes closer to the disk ($|b| \alt 20^\circ$), the
spectrum of the emission correlated with the bubbles possesses a
pronounced spectral feature in $E_\gamma^2d\Phi_\gamma/dE_\gamma$ peaking at
$E_\gamma \sim 1-4~{\rm GeV}$, which cannot be produced by any
realistic spectrum of electrons~\cite{Hooper:2013rwa}. This implies
that a second (non-inverse-Compton) emission mechanism must be
responsible for the bulk of the low-energy, low-latitude emission. The
spectral shape of this second component is similar to the one reported
from the GC.  The intrinsic non-inverse-Compton emission appears
spatially consistent with a luminosity per volume falling
approximately as $r^{-2.4} - r^{-2.8}$.  As a consequence, the
spectral feature visible in the low-latitude bubbles is most likely
the extended counterpart of the GC excess, now detected out to at
least $r \sim 2-3~{\rm kpc}$. Even though millisecond pulsars possess
a spectral cutoff at approximately the required energy, these sources
exhibit a spectral shape that is much too soft at sub-GeV energies to
accommodate this signal~\cite{Hooper:2013nhl}.

The spectrum and angular distribution of the signal is broadly
consistent with one predicted from $\sim 10~{\rm GeV}$ dark matter
particles annihilating to leptons, or from $\sim 50~{\rm GeV}$ dark
matter particles annihilating to quarks, following a distribution
similar to, but slightly steeper than, the canonical
Navarro--Frenk--White (NFW) profile. In either case, the morphology of the
$\gamma$-ray signal requires a dark matter distribution that scales
approximately as $\rho_{\rm DM} \propto r^{-1.2} - r^{-1.4}$; that is, 
the annihilation rate per volume is proportional to the square of the
dark matter density. Such a dark matter distribution is in good 
agreement with current observational constraints~\cite{Iocco:2011jz}.

For the 10~GeV dark matter candidate, the normalization of the
observed signal requires a velocity average annihilation cross section
on the order of
\begin{equation}
\langle \sigma_\tau v \rangle
\sim 2 \times 10^{-27}~{\rm cm}^{3}/{\rm s} = 1.7 \times 10^{-10}~{\rm GeV}^{-2} \,,
\label{fermibubble}
\end{equation}
up to overall uncertainties in the normalization of the halo
profile~\cite{Hooper:2012ft}. This light mass scenario has been
further invigorated by various observations reported by the
DAMA/LIBRA~\cite{Bernabei:2010mq},
CoGeNT~\cite{Aalseth:2010vx,Aalseth:2011wp},
CRESST~\cite{Angloher:2011uu}, and CDMS~\cite{Agnese:2013rvf}
collaborations, each of which report signals consistent with a dark
matter particle of similar mass. These four experiments make use of
different technologies, target materials, and detection strategies,
but each reports results that are not compatible with known
backgrounds but which can be accommodated by a light dark matter
particle with a mass of about 10 GeV and an elastic scattering cross
section with nucleons of $1 - 2 \times 10^{-41}~{\rm
  cm}^2$~\cite{Fitzpatrick:2010em,Chang:2010yk,Hooper:2010uy,Buckley:2010ve}.

Dark matter particles can elastically scatter with nuclei in the Sun,
leading to their gravitational capture and subsequent
annihilation. Electrons and muons produced in such annihilations
quickly lose their energy to the solar medium and produce no
observable effects. Annihilations to taus, on the other hand, produce
neutrinos which, for a 10~GeV, can be observed by
Super-Kamiokande. For the required branching into $\tau^+ \tau^-$ of
about 10\% -- as given by (\ref{fermibubble}) -- existing data
constrain the dark matter spin-independent elastic scattering cross
section with protons to be less than $4 \times 10^{-41}~{\rm
  cm}^{2}$~\cite{Hooper:2008cf,Kappl:2011kz}.

For the 50~GeV dark matter particle, the normalization of the
observed signal requires a velocity average annihilation cross section
on the order of
\begin{equation}
\langle \sigma_b v \rangle
\sim 8 \times 10^{-27}~{\rm cm}^{3}/{\rm s} = 6.7 \times 10^{-10}~{\rm GeV}^{-2} \, .
\label{fermibubble2}
\end{equation}
The XENON-100 Collaboration reported a 90\%~C.L. bound on the elastic
scattering cross section with nuclei of ${\cal O} (10^{-44}~{\rm
  cm}^2)$~\cite{Aprile:2011hi}. A later analysis arrived at
alternative conclusions allowing for a signal of two events with a
favored mass of 12~GeV and large error contour extending to about 50~GeV~\cite{Hooper:2013cwa}.

It is worthwhile to point out that the bounds from the combined
analysis of 10 dwarf
spheroidals~\cite{Ackermann:2011wa,GeringerSameth:2011iw}, galaxy
clusters~\cite{Ackermann:2010rg}, or diffuse $\gamma$-ray
emission~\cite{Abazajian:2010sq,Abdo:2010dk} are not sensitive enough to
probe the velocity average annihilation cross sections
(\ref{fermibubble}) and (\ref{fermibubble2}).

In this paper we study the minimal hidden sector of Weinberg's Higgs
portal model~\cite{Weinberg:2013kea}, and we examine to what extent
its free parameters can be adjusted to explain the low-latitude
$\gamma$-ray emission from the \emph{Fermi} bubbles. The layout of the
paper is as follows. In Sec.~II we outline the basic setting of the
model. In Sec.~III we review the constraints related to experimental
searches for new physics at the LHC. After that, in Sec.~IV we turn
our attention to the prospects for direct dark matter searches. In
Sec.~V we study the constraints from cosmological observations. In
Sec.~VI we present the main results of this work. We begin by
constraining the parameter space in the Higgs sector along a
correlation of the \emph{Fermi} bubbles' $\gamma$-ray signal with the
dark matter annihilation cross section into SM fermions. We then
further constrain the parameter space by matching the thermal relic
abundance of dark matter with the value inferred by cosmological
observations. Lastly, in Sec.~VII we explore a region of parameter
space which cannot accommodate \emph{Fermi} observations but remains
interesting in itself. In Sec.~VIII we summarize our findings.

\section{$\bm{W}$-WIMP\lowercase{s}}

Weinberg's Higgs portal model is based on a broken global $U(1)$
symmetry associated with the dark matter charge $W$: the number of
weakly interacting massive particles (WIMPs) minus the number of their
antiparticles.  The  hidden sector contains a
Dirac field $\psi$ (carrying WIMP quantum number $W = +1$) and a
complex scalar field (with $W = 2$, so that its expectation value
leaves an unbroken reflection symmetry $\psi \to - \psi$). All SM
fields are assumed to have $W=0$. 

The scalar potential consists of the SM component $[s]$, the
isomorphic component in the hidden sector $[h]$, and the quartic
interaction coupling between the two sectors with strength
$\eta_\chi$. The Lagrangian density for the scalar sector reads
\beqa 
\mathscr{L} = | \pd \Phi_h |^2 + | \pd \Phi_s |^2 +
\mu_h^2 |\Phi_h|^2 - \lambda_h |\Phi_h|^4 +
\mu_s^2 |\Phi_s|^2 - \lambda_s |\Phi_s|^4 -
\eta_\chi |\Phi_h|^2|\Phi_s|^2  \,,
\label{eq:one}
\eeqa 
where $\Phi_s$ is the SM scalar doublet and $\Phi_h$ is
a complex scalar field. We separate a massless Goldstone boson field
$\alpha(x)$ and a massive radial field $r(x)$ by defining
\begin{equation}
\Phi_h(x) =  \frac{1}{\sqrt{2}} \ r(x) \ e^{i\, 2 \alpha(x)} \,, 
\end{equation}
where $r(x)$ and $\alpha(x)$ are real, with the phase of $\Phi_h(x)$
adjusted to make the vacuum expectation value (VEV) of $\alpha (x)$
zero. The $SU(2) \times U(1)$ symmetry of the SM is (of course) broken
by a nonvanishing VEV of the neutral component $\phi$
of the scalar doublet,
\begin{equation}
\Phi_s = \frac{1}{\sqrt{2}} \ \left( \begin{array}{c} G^\pm \\
v_\phi + \phi' + iG^0 \end{array} \right) ,
\end{equation}
where $v_\phi \simeq 246~{\rm GeV}$. The $G$ fields are the familiar Goldstone
bosons, which are eaten by the vector bosons ({\it i.e.} the $G^\pm$
become the longitudinal components of the charged $W$ boson and $G^0$
becomes the longitudinal component of the $Z$ boson). In terms of real
fields the Lagrangian density (\ref{eq:one}) takes the form
\begin{equation}
\mathscr{L} = \frac{1}{2} \pd r^2 +  \frac{1}{2} \pd \phi^2 + 2 r^2 \pd \alpha^2 +
\frac{\mu_h^2}{2} r^2 - \frac{\lambda_h}{4}r^4 + \mu_s^2
|\phi|^2 - \lambda_s |\phi|^4 - \frac{\eta_\chi}{2} r^2 |\phi|^2
 \, .
\label{eq:three}
\end{equation}
The $U(1)$ symmetry of $W$ conservation is also broken and $r$ gets a VEV
 \begin{equation}
r (x)  = v_r  + r' (x) \,,  
\label{eq:four}
\end{equation}
with $v_r$ real and non-negative.

We demand the scalar
potential obtains its minimum value at
\beq
{\cal V} =   -\frac{\mu_h^2}{2} v_r^2 + \frac{\lambda_h}{4} v_r^4 - \frac{\mu_s^2}{2}  v_\phi ^2
+ \frac{\lambda_s}{4} v_\phi^4 + \frac{\eta_\chi}{4} v_r^2 v_\phi^2 \, . 
\label{eq:five}
\eeq
Physically, the most interesting solutions to the minimization of
(\ref{eq:five}), 
\begin{equation}
\pd_{v_r} {\cal V} = - \mu_h^2 v_r +  \lambda_h v_r^3 +  \frac{\eta_\chi}{2} v_r  v_\phi^2 =0 
\end{equation}
 and
\begin{equation}
\pd_{v_\phi} {\cal V}  =  - \mu_s^2 v_\phi  +  \lambda_s v_\phi^3 
+ \frac{\eta_\chi}{2}  v_r ^2  v_\phi =0 \,, 
\end{equation}
are obtained for $v_r $ and $v_\phi$ both nonvanishing
\begin{equation}
v_\phi^2  = \frac{1}{\lambda_s}\left(\mu_s^2 - \frac{\eta_\chi v_r^2}{2}  \right)
\end{equation}
and
\begin{equation}
v_r^2  =
\frac{1}{\lambda_h}\left(\mu_h^2 - \frac{\eta_\chi v_\phi^2}{2} \right) \,,
\label{eq:six}
\end{equation}
respectively. To compute the scalar masses, we must expand the
potential  around the minima 
 \beqa 
 \mathscr{L} & = &\frac{1}{2}
(\pd r')^2 \nonumber
\\
& + & 2 v_r^2 \pd \alpha^2 + 4 v_r r' \pd \alpha^2 +  2 r'^2 \pd \alpha^2 \nonumber
\\
& - & \lambda_h v_r^2 r'^2 - \lambda_s v_\phi^2 \phi^{'2} -
\eta_\chi v_r v_\phi r' \phi' + \cdots \,, 
\label{eq:seven}
\eeqa 
where the dots indicate 3-point and 4-point interactions, as well as
the SM interactions.  There is a mixing term present for $r'$ and
$\phi'$. We find the fields of definite mass by diagonalizing the mass
matrix for $r'$ and $\phi'$.  We denote by $H$ and $h$ the scalar
fields of definite masses, $m_H = 125~{\rm GeV}$ and $m_h$, respectively. After a
bit of algebra, the explicit expressions for the scalar mass
eigenvalues and eigenvectors are given by 
\begin{equation}
m^2_h =  \lambda _h v_r^2 +  \lambda_s v_\phi^2 - \sqrt{(\lambda_s v_\phi^2 - \lambda _h
  v_r^2)^2 + (\eta_\chi v_r  v_\phi )^2} 
\end{equation}
and
\begin{equation}
m^2_H =  \lambda _h v_r^2 + \lambda _s v_\phi^2 +\sqrt{(\lambda_s v_\phi^2 - \lambda _h
  v_r^2)^2 + (\eta_\chi v_r  v_\phi )^2} \,,
\label{eq:eight}
\end{equation}
with
\begin{equation}
\left( \begin{array}{c} h \\ H \end{array}\right) =
\left( \begin{array}{cc} \cos{\chi}& -\sin{\chi}\\ \sin{\chi}& \phantom{-}\cos{\chi}
\end{array}\right) \left( \begin{array}{c} r' \\ \phi' \end{array}\right) \, ,
\label{eq:nine}
\end{equation}
where $\chi \in [-\pi/2,\pi/2]$ also fulfills
\begin{equation}
\sin 2\chi = \frac{ \eta_\chi v_\phi v_r}{\sqrt{(\lambda_s v_\phi^2 - \lambda _h
  v_r^2)^2 + (\eta_\chi v_r  v_\phi )^2} } = \frac{2 \eta_\chi v_\phi v_r}{m_H^2 - m_h^2},
\label{correa}
\end{equation}
and
\begin{equation}
\cos 2\chi = \frac{ \lambda _s  v_\phi^2 - \lambda_h v_r^2}{\sqrt{(\lambda_s v_\phi^2 - \lambda _h
  v_r^2)^2 + (\eta_\chi v_r  v_\phi )^2} } \ , 
\end{equation}
yielding
\begin{equation}
\tan 2\chi = \frac{ \eta_\chi v_r  v_\phi}{ \lambda_s v_\phi^2 -  \lambda_h v_r^2} \, .
\label{eq:ten}
\end{equation}
The Goldstone boson in (\ref{eq:seven}) has to be be renormalized so
that it resumes the standard canonical form. This is achieved through
scaling $\alpha \rightarrow \alpha' = 2 v_r \alpha$, videlicet,
\beq
2 v_r^2 \pd \alpha^2 + 4 v_r r' \pd \alpha^2 +  2 r'^2 \pd \alpha^2 \rightarrow \frac{1}{2} \pd \alpha'^2 + \frac{1}{v_r} r' \pd \alpha'^2 + \frac{1}{2 v_r^2} r'^2 \pd \alpha'^2.
\label{eq:eleven} 
\eeq 

Adding in the dark matter sector requires at least one Dirac field
\beq
\mathscr{L}_\psi = i \bar{\psi}\gamma \cdot \pd \psi - m_\psi \bar{\psi} \psi 
- \frac{f}{\sqrt{2}} \bar{\psi^c} \psi \Phi_h^\dagger  - \frac{f^*}{\sqrt{2}} \bar{\psi} \psi^c \Phi_h  .
\label{eq:fortyone}
\eeq We assign $\psi$ a charge $W = 1$, so that the Lagrangian is
invariant under the global transformation $e^{i W \alpha}$. Treating 
the transformation as local allows us to express $\psi$ as \beq
\psi(x) = \psi'(x) e^{i \alpha(x)}.
\label{eq:fortytwo}
\eeq
We can now rewrite (\ref{eq:fortyone}) in terms of $\psi',$ $\alpha,$ and $r$
\beq
\mathscr{L}_\psi = i \bar{\psi}'\gamma \cdot \pd \psi' - (\bar{\psi}' \gamma \psi' )  \cdot \pd \alpha - m_\psi \bar{\psi}' \psi' - \frac{f}{2} \bar{\psi'}^{c} \psi' r  - \frac{f}{2}  \bar{\psi}' {{\psi'}^{c}} r  \,,
\label{eq:fortythree}
\eeq where we have taken $f$ to be real. Once $r$ achieves a VEV we
can expand the dark matter sector to get \beqa \mathscr{L}_\psi &=&
\frac{i}{2}\left(\bar{\psi}'\gamma \cdot \pd \psi' + \bar{\psi'}^{c}
  \gamma \cdot \pd \psi^{c'} \right), \nonumber
\\
&-& \frac{m_\psi}{2} \left( \bar{\psi}' \psi' + \bar{\psi'}^{c}
  {\psi'}^{c} \right)-\frac{f v_r}{2} \bar{\psi'}^{c} \psi' - \frac{f
  v_r}{2} \bar{\psi}' {\psi'}^{c} , \nonumber
\\
&-& \frac{1}{2} (\bar{\psi}' \gamma \psi' - \bar{\psi'}^{c} \gamma
{\psi '}^{c} ) \cdot \pd \alpha , \nonumber
\\
&-& \frac{f}{2} r' \left( \bar{\psi'}^{c}\psi' + \bar{\psi}' {\psi'}^{c}
\right).
\label{eq:fortyfour}
\eeqa
Note that we have made the Lagrangian explicitly symmetric via relations like
\begin{eqnarray}
\psi^c &=& C \bar{\psi}^T  \\
\label{eq:fourtyfive}
\bar{\psi^c} \psi^c &=& ( - \psi^T C^{-1} C \bar{\psi}^T) = \bar{\psi} \psi \\
\bar{\psi}_c \gamma \cdot \pd \psi_c &=& - \psi^T C^{-1} \gamma C \cdot \pd \bar{\psi}^T \nonumber
\\
\label{eq:fourtysix}
&=& \psi^T \gamma^T \cdot \pd \bar{\psi}^T = - (\pd \bar{\psi} \cdot \gamma \psi) \rightarrow \bar{\psi} \gamma \cdot \pd \psi \, .
\end{eqnarray} 
In (\ref{eq:fourtyfive}) we used the Grassman nature of the spinor
fields; in the second line of (\ref{eq:fourtysix}) we used integration
by parts to transfer the derivative onto the $\psi$ field. Similar
results can be found for the other expressions.

Diagonalization of the $\psi'$ mass matrix  generates the mass eigenvalues,
\begin{equation}
m_\pm =  m_\psi \pm  f  v_r, 
\end{equation}
for the two mass eigenstates
\begin{equation}
\psi_- = \frac{i}{\sqrt{2}} \lpa \psi'^c - \psi'  \rpa
  \quad {\rm and}   \quad \psi_+ = \frac{1}{\sqrt{2}}\lpa \psi'^c+\psi' \rpa  \, .
\label{eq:fourtyseven}
\end{equation}
In this basis, the act of charge conjugation on $\psi_\pm$ results in
\beq
\psi^c_\pm =  \psi_\pm.
\label{eq:fourtyeight}
\eeq This tells us that the fields $\psi_\pm$ are Majorana
fermions. The Lagrangian is found to be 
\beqa \mathscr{L}_\psi
&=&\frac{i}{2}\bar{\psi_+}\gamma \cdot \pd \psi_+ +
\frac{i}{2}\bar{\psi_-}\gamma \cdot \pd \psi_- - \frac{1}{2} m_+
\bar{\psi}_+ \psi_+ - \frac{1}{2}m_- \bar{\psi}_- \psi_- , \nonumber
\\
&-&\frac{i}{4 v_r} (\bar{\psi}_+ \gamma \psi_- - \bar{\psi}_-
\gamma \psi_+) \cdot \pd \alpha' , \nonumber
\\
& -& \frac{f}{2} r' (\bar{\psi}_+\psi_+ - \bar{\psi}_- \psi_-).
\label{eq:fortynine} 
\eeqa
We must now put $r'$ into its massive field representation, for which the interactions of interest are
\beq
-\frac{f \sin \chi}{2} H (\bar{\psi}_+\psi_+ -  \bar{\psi}_-
\psi_-) - \frac{f  \cos \chi}{2} h (\bar{\psi}_+\psi_+ -  \bar{\psi}_- \psi_-). 
\label{eq:fifty}
\eeq This leads to 3-point interactions between the $W$-WIMPs and the
Higgs boson of the SM.

In summary, instead of one Dirac $W$-WIMP, there are two Majorana
$W$-WIMPs of different masses. However, the heavier $W$-WIMP will
decay into the lighter one by emitting a Goldstone boson, while the
lighter one is kept stable by an unbroken reflection
symmetry. Therefore in this model we can expect that the Universe
today will contain only one type of Majorana $W$-WIMP, the lighter one
$w$, with mass $m_w$ equal to the smaller of $m_\pm$. Throughout,
$\Delta m = |m_+ - m_-| = 2 | f v_r|$ denotes the mass splitting of
the $W$-WIMP states.  (The most common variables used in this article
are summarized in Table~\ref{table:0}.)

A cautionary note is worth taking on board at this juncture. It has
long been known that the spontaneous breaking of a global $U(1)$
symmetry has several disconnected and degenerate vacua (the phase of
the vacuum expectation value $\langle 0 | \Phi_h |0 \rangle$ can be
different in different regions of space, and actually we expect it to
be different in casually disconnected regions), leading to
catastrophic domain-wall structures in the early
Universe~\cite{Sikivie:1982qv,Vilenkin:1982ks}. In the spirit
of~\cite{Sikivie:1982qv}, it may be possible to introduce a small
explicit breaking of the symmetry, such that the domain walls
disappear before dominating the matter density of the Universe, while
leaving \mbox{(pseudo-)Goldstone} bosons and the same dark matter
phenomenology.\footnote{Other approaches, if exceedingly fine-tuned,
  may offer alternative
  solutions~\cite{Hill:1989sd,Linde:1990yj,Turner:1990uz,Dvali:1991ka,Hiramatsu:2012sc}.}

The absence of new physics signals at the LHC place constraints on the
model. We discuss this next.

\begin{table}
\caption{Definition of most common variables. \label{table:0}}
\begin{tabular}{ll}
\hline
\hline
$\Phi_s$  & ~~SM scalar doublet \\
\hline
$\Phi_h$  & ~~Complex scalar field \\
\hline
$\phi$ & ~~Neutral component of the scalar doublet \\
\hline
$r$ & ~~Massive radial field \\
\hline
$\alpha$ & ~~Goldstone boson \\
\hline
$v_\phi$ & ~~Vacuum expectation value of $\phi$ \\
\hline
$v_r$ & ~~Vacuum expectation value of $r$ \\
\hline
$H$ & ~~SM Higgs boson \\
\hline
$h$ & ~~Hidden scalar \\
\hline
$\eta_\chi$ & ~~Quartic interaction coupling between SM and hidden sectors  \\
\hline
$\chi$ & ~~$H$-$h$ mixing angle \\
\hline
$w$ & ~~Lightest $W$-WIMP \\
\hline
$\Delta m$ & ~~$W$-WIMP mass splitting\\
\hline
$f$ & ~~Coupling between hidden Majorana fermions and complex scalar field\\
\hline
\hline
\end{tabular}
\end{table}

\section{Constraints from Collider Experiments}

The recent discovery~\cite{Aad:2012tfa,Chatrchyan:2012ufa} of a new
particle with properties consistent with the SM Higgs boson is without
any doubt the most compelling news from the LHC. With the measurements
in various channels, a comprehensive study of the properties of the
Higgs-like state becomes possible and has the potential for revealing
whether or not the Higgs sector is as simple as envisioned in the
SM. Since invisible decays reduce the branching fraction to the
(visible) SM final states, it is to be expected that ${\cal B} (H \to
\, {\rm invisible})$ is strongly constrained. Indeed ${\cal B} (H \to
\, {\rm invisible})$ is known to be less than about 19\% at
95\%C.L.~\cite{Espinosa:2012vu,Cheung:2013kla,Giardino:2013bma,Ellis:2013lra}. Thus,
the mixing of the SM with the hidden sector must be weak. Note that
for $\eta_\chi \ll 1$  the relations between masses and angles then 
become
\begin{equation}
m^2_h \approx 2 \lambda _h v_r^2, \quad \quad m^2_H \approx  2 \lambda_s
v_\phi^2, \quad  \quad
\tan 2\chi \approx  \frac{2 \eta_\chi v_r  v_\phi}{m_H^2 - m_h^2} \, ,
\label{eq:ten-1}
\end{equation}
where we have assumed $ \lambda _s v_\phi^2 > \lambda _h v_r^2$. For a
Higgs width of about 4~MeV, the partial width for decay into 
unobserved particles is found to be
\begin{equation}
\Gamma_{H \to \, {\rm invisible}} < 0.8~{\rm MeV} \, . 
\label{LHCwidth}
\end{equation}

The phenomenology of a Higgs portal to the hidden sector depends on
whether the SM Higgs particle is lighter or heavier than the new
companion. In this study we take $m_H > m_h$. The decay rate into
invisible stuff, $\Gamma_{H \to \, {\rm invisible}}$, has two distinct
contributions: $\Gamma_{H \to \, {\rm invisible}}^{\rm SM}$ and
$\Gamma_{H \to \, {\rm hidden}}$. The former is dominated by $H \to 2
Z \to 4 \nu$, with an invisible $Z$ branching ratio of 4\%. The $4
\nu$ rate can be predicted from observed decays $H \to 2Z \to 4l$. For
the sake of simplicity, hereafter we will omit the contribution of
$\Gamma_{H \to \, {\rm invisible}}^{\rm SM}$. Unless expressly stated
otherwise herein, we assume $m_w + \Delta m > m_H/2$ and thus $H$
decays (invisibly) into the hidden sector via three channels: $H \to 2
\alpha'$, ${H \to 2 h}$, and $H \to 2 w$. From the event rates for
visible Higgs production and decay channels, we could derive upper
bounds on non-SM admixtures in the wave function of the Higgs boson
and on the new three invisible decay channels. To this end we now
compute the decay rates for these three processes.

\subsection{$\bm{\Gamma_{H\to 2 \alpha'}}$}

Substituting in (\ref{eq:eleven}) $r'$ by the field of definite mass,
$r' = h \cos \chi + H \sin \chi$, we can write the Higgs--Goldstone
boson interaction term as \beq \frac{1}{v_r} r' \pd \alpha'^2
\rightarrow \frac{\sin \chi}{v_r} H (\pd \alpha')^2 + \frac{\cos
  \chi}{v_r} h (\pd \alpha')^2 \, .
\label{eq:twelve}
\eeq
Using (\ref{eq:twelve}) we write the Feynman rule for
        interactions of the type $H, \ \alpha', \ \alpha'$ as 
\beq
-i\frac{2 \sin \chi}{v_r} k \cdot k' ,
\label{eq:thirteen}
\eeq where $k \, (k')$ is the 4-momentum of the incoming (outgoing)
$\alpha'$ particle, and the factor of $2$ is a symmetry factor, as one
can exchange incoming-outgoing $\alpha'$ twice.  From this 3-point
interaction we can calculate the decay width of the SM Higgs $H$
into 2 Goldstone bosons $\alpha'$. In the rest frame of the Higgs, the
differential decay probability per unit time is given by 
\beq
d\Gamma_{H \rightarrow 2 \alpha'} = \frac{1}{2 m_H} \left(\frac{2 \sin \chi}{v_r} k_1 \cdot k_2\right)^2 \, d\mathcal{Q} \,,
\label{eq:fourteenW}
\eeq
where 
\beqa
d\mathcal{Q} &=& \frac{1}{2!} \ \frac{d^3 k_1}{(2\pi)^3 2 k_1}\frac{d^3 k_2}{(2\pi)^3 2 k_2}
(2\pi)^4 \delta(m_H - k_1 - k_2)\delta^{(3)}(\textbf{k}_1 +
\textbf{k}_2)  \nonumber
\\
&=& \left. \frac{1}{16} \frac{d\Omega_{k_1}}{(2 \pi)^2} \right|_{k_1 = m_H/2} 
\eeqa
is the phase space for a two-body final state (the factor of $1/2!$
is included because of  identical particles in the final state).  After some
algebra (\ref{eq:fourteenW}) can be rewritten as
\begin{equation}
d\Gamma_{H \rightarrow 2 \alpha'}  =  \frac{d \Omega_{k_1}}{128~\pi^2
  m_H} \left[\frac{2 \, \sin\chi}{v_r} \, 2 \, \left( \frac{m_H}{2} \right)^2 \right]^2 \, .
\end{equation}
The partial decay width can now be expressed as
\begin{equation}
\Gamma_{H \rightarrow 2 \alpha'}  = \frac{1}{32 \pi}
\left(\frac{\sin \chi}{v_r}\right)^2 m_H^3 \, .
\end{equation}
For $m_H \gg m_h$ and $m_H^2 \gg 2 \, \eta_\chi \, v_r  v_\phi$,
we can use the small angle approximation
\begin{equation}
\sin \chi \approx \chi =  \eta_\chi \, v_r v_\phi
/(m_H^2-m_h^2) \, . 
\end{equation}
In this very good approximation
the decay width becomes
\beq
\Gamma_{H \rightarrow 2 \alpha'} =\frac{1}{32 \pi}
\left(\frac{\eta_\chi \, v_\phi }{m_H^2-m_h^2}\right)^2 m_H^3.
\label{eq:fifteen}
\eeq

\subsection{$\bm{\Gamma_{H\to 2 h}}$}

We begin by expanding the scalar potential  $\CV$ around the VEVs of $r$
and $\phi$ after which we diagonalize the mass matrix. Together this
requires that we expand around the fields \beqa r(x) &=& v_r + h \cos
\chi + H \sin \chi \ , \nonumber
\\
\phi(x) &=& v_\phi + H \cos \chi - h \sin \chi \ , \eeqa 
 which puts $\CV$  in the form 
\begin{eqnarray}
\CV &=& \frac{1}{2}
m_H^2 H^2 + \frac{1}{2} m_h^2 h^2 \nonumber \\
& - &  \frac{1}{16} \lpa \eta_\chi + 3 (\lambda_h+\lambda_s)+3(\eta_\chi
-\lambda_h-\lambda_s)\cos 4\chi \rpa H^2 h^2 \nonumber
\\
&- &\frac{1}{4}   v_\phi \cos \chi [6 \lambda_s - \eta_\chi +
3(\eta_\chi - 2 \lambda_s)\cos 2\chi]  H h^2  \nonumber \\
& -& \frac{1}{4} v_r \sin \chi [6 \lambda_h-\eta_\chi  -
3 (\eta_\chi - 2 \lambda_h) \cos 2 \chi ]  H h^2 \nonumber
\\
&- &\frac{1}{4} \lpa \lambda_s \cos^4 \chi + \eta_\chi \cos^2 \chi
\sin^2 \chi + \lambda_h \sin^4 \chi \rpa H^4 \nonumber
\\
&- &\frac{1}{4} \lpa \lambda_h \cos^4 \chi + \eta_\chi \cos^2 \chi
\sin^2 \chi + \lambda_s \sin^4 \chi \rpa h^4 \nonumber
\\
&+&\frac{1}{2} \lpa v_\phi \sin \chi ( \eta_\chi \cos^2 \chi + 2
\lambda_s \sin^2 \chi) - v_r (2 \lambda_h \cos^3 \chi + \eta_\chi \cos
\chi \sin^2 \chi) \rpa h^3 \nonumber
\\
&- &\frac{1}{2} \lpa v_r \sin \chi (\eta_\chi \cos^2 \chi + 2 \lambda_h
\sin^2 \chi) + v_\phi (2\lambda_s \cos^3 \chi + \eta_\chi \cos \chi
\sin^2 \chi)\rpa H^3 \nonumber
\\
&- &\frac{1}{4} \lpa \lambda_h - \lambda_s +(\lambda_s + \lambda_h -
\eta_\chi) \cos 2\chi \rpa \sin 2\chi \ H h^3 \nonumber
\\
&+ & \frac{1}{4} \lpa \lambda_s - \lambda_h +(\lambda_s + \lambda_h -
\eta_\chi) \cos 2\chi \rpa \sin 2\chi \ H^3 h \nonumber
\\
&+ &\frac{1}{2}  v_\phi \sin \chi [2 (3 \lambda_s - \eta_\chi)
\cos^2 \chi + \eta_\chi \sin^2 \chi] H^2 h  \nonumber \\
&+ & \frac{1}{2}v_r (\eta_\chi \sin \chi \sin 2
\chi - 6 \lambda_h \cos \chi \sin^2 \chi - \eta_\chi \cos^3 \chi) 
H^2 h \, .
\end{eqnarray} 
Since $\chi < 1$  we first
expand the potential  around $\chi=0$, and then using (\ref{correa})  we further expand around
$\eta_\chi = 0$  retaining only the terms first order in
$\eta_\chi$;  this results in \beqa \CV &\approx& \frac{1}{2} m_H^2 H^2
+ \frac{1}{2} m_h^2 h^2 \nonumber
\\
&-&\frac{\eta_\chi }{4} H^2 h^2 - \frac{\lambda_h}{4} h^4 -
\frac{\lambda_s}{4} H^4 - \frac{\eta_\chi \lambda_h v_r v_\phi}{m_H^2
  -m_h^2} H h^3 + \frac{\eta_\chi \lambda_s v_r v_\phi}{m_H^2 -m_h^2}
H^3 h \nonumber
\\
&-& \lambda_h v_r h^3 - \lambda_s v_\phi H^3 -\frac{\eta_\chi v_\phi}{2}
\lpa \frac{6 \lambda_h v_r^2}{m_H^2 - m_h^2} +1 \rpa H h^2 +
\frac{\eta_\chi v_r}{2} \lpa \frac{6 \lambda_s v_\phi^2}{m_H^2 -
  m_h^2} -1\rpa H^2 h \ . \nonumber 
\eeqa 
Using (\ref{eq:ten-1}) we can manipulate this expression to write the scalar
potential as 
\beqa \CV &\approx&\frac{1}{2} m_H^2 H^2 + \frac{1}{2}
m_h^2 h^2 \nonumber
\\
&-&\frac{\eta_\chi }{4} H^2 h^2 - \frac{\lambda_h}{4} h^4 -
\frac{\lambda_s}{4} H^4 - \frac{\eta_\chi \lambda_h v_r v_\phi}{m_H^2
  -m_h^2} H h^3 + \frac{\eta_\chi \lambda_s v_r v_\phi}{m_H^2 -m_h^2}
H^3 h \nonumber
\\
&-& \frac{m_h^2}{2 v_r} h^3 - \frac{m_h^2}{2 v_\phi}H^3 -\frac{\eta_\chi
  v_\phi}{2} \lpa \frac{m_H^2 + 2 m_h^2}{m_H^2 - m_h^2} \rpa H h^2 +
\frac{\eta_\chi v_r}{2} \lpa \frac{2 m_H^2 + m_h^2}{m_H^2 - m_h^2}
\rpa H^2 h \ . 
\eeqa
Under the approximations
taken previously, $m_H \gg m_h$ and $m_H^2 \gg 2 \eta_\chi v_r v_\phi$,
the relevant $Hhh$ interaction term results in 
\beq -\frac{\eta_\chi v_\phi}{2} \lpa \frac{m_H^2 + 2
  m_h^2}{m_H^2-m_h^2} \rpa H h^2. \eeq
The differential decay probability  per unit time is given by
\begin{equation}
d\Gamma_{H \to 2h}  =  \frac{1}{2 m_H} \left( \frac{\eta_\chi
  v_\phi}{m_H^2 - m_h^2} \right)^2 \ \left(m_H^2 + 2 m_h^2 \right)^2
\frac{1}{2!} \frac{k^2 \, dk \, d \Omega}{(2 \pi)^2 4  E_k}
\frac{1}{2k} \ \delta \left( m_h - 2 \sqrt{k^2 + m_H^2} \right) \, .
\end{equation}
The partial $H \to 2 h$ decay width can now be expressed as
\begin{equation}
\Gamma_{H \to 2h} =  \frac{1}{32 \, \pi \, m_H^2}  \left( \frac{\eta_\chi
  v_\phi}{m_H^2 - m_h^2} \right)^2 \  \left(m_H^2 + 2 m_h^2 \right)^2
\ \sqrt{m_H^2 - 4 m_h^2} \, .
\end{equation}
In the limit $m_H \gg m_h$ we obtain
\begin{equation}
\Gamma_{H \to 2h} = \frac{1}{32 \pi} \left( \frac{\eta_\chi
  v_\phi}{m_H^2 - m_h^2} \right)^2  \ m_H^3 \, .
\end{equation}
 
\subsection{$\bm{\Gamma_{H\to 2 w}}$}

For $m_w < m_H/2$, the $r-\phi$ mixing allows the Higgs boson to decay
into pairs of the lightest $W$-WIMP.  We obtain the invariant
amplitude for this process (a description of Feynman rules for
Majorana fermions can be found in Ref.~\cite{Srednicki:2007qs}),
\begin{equation} 
i \CM = i f \sin \chi \bar{u}(p) v(p') \,, 
\end{equation}
where $u(p)$ and $v(p)$ are Dirac spinors. The spin average rate is
given by
\begin{equation}
\sum_s | \CM|^2 = 4 f^2 \sin^2 \chi( p\cdot p' - m_w^2).  
\end{equation}
The
partial $H$-decay rate into $2w$ is
\begin{eqnarray}
d\Gamma_{H \to 2 w} &=& \frac{|\CM|^2}{2 m_H}\frac{d^3 p'}{(2\pi)^3 2 E_{p'}} \frac{d^3 p}{(2\pi)^3 2 E_p} (2\pi)^4 \delta^{(3)}(\textbf{p}'+\textbf{p}) \delta(m_H - p' - p) \nonumber
\\
&=& \frac{1}{2!} \frac{d\Omega}{64 \pi^2 m_H^2} \sqrt{m_H^2-4m_w^2} |\CM|_{\textbf{p}'=-\textbf{p}, \ p = \sqrt{(m_H/2)^2-m_w^2}}^2 \, ,
\end{eqnarray}
and so the partial width for this decay is given by
\beq
\Gamma_{H \to 2 w} = \frac{ 2( m_H^2- 4 m_w^2)  }{32 \pi m_H^2} \left
  ( \frac{f \eta_\chi v_r v_\phi}{m_H^2-m_h^2} \right)^2
\sqrt{m_H^2-4m_w^2} \, .
\label{eq:fiftyone}
\eeq
For  $m_H \gg 2 m_w$, (\ref{eq:fiftyone}) becomes 
\beq
\Gamma_{H \rightarrow 2 w} = \frac{1}{16 \pi} \left( \frac{f
    \eta_\chi v_r   v_\phi }{m_H^2 - m_h^2}\right)^2
\sqrt{m_H^2-4m_w^2}.
\label{godoy}
\eeq

\subsection{$\bm{\Gamma_{H\to \ {\rm hidden}}}$}

All in all, the decay width of the Higgs into the
hidden sector  is given by \beq \Gamma_{H
  \rightarrow {\rm hidden}}  =  \frac{1}{16 \pi}
\left(\frac{\eta_\chi \, v_\phi }{m_H^2-m_h^2}\right)^2 m_H^3 +
\frac{1}{16 \pi} \left( \frac{f \, \eta_\chi \ v_r \
 \ v_\phi }{m_H^2 - m_h^2}\right)^2
\sqrt{m_H^2-4m_w^2}.
\label{eq:invis}
\eeq
Assuming $m_H \gg m_h$, this decay width is 
\beq
\Gamma_{H \rightarrow {\rm hidden}}  = \frac{\eta_\chi^2
  v_\phi^2}{16 \pi m_H } + \frac{\eta_\chi^2 \Delta m^2 v_\phi^2}{64
  \pi m_H^3} .
\label{eq:fifteen-1}
\eeq 
Comparing (\ref{LHCwidth}) and (\ref{eq:fifteen-1}), we obtain
\begin{equation}
|\Delta m| \alt  2 m_H \ \sqrt{ \frac{8.3 \times 10^{-5}}{\eta_\chi^2 } - 1} \, ,
 \label{eq:DeltaMlim}
\end{equation}
which is satisfied if $|\eta_\chi| < 0.009$.

\section{Constraints from Direct Detection Experiments}

Direct detection experiments attempt to observe the recoil from the
elastic scattering of dark matter particles interacting with nuclei in
the detector. Since the late 1990's the \mbox{DAMA/NaI}
Collaboration~\cite{Bernabei:1998fta} has been claiming to observe the
expected annual modulation of the dark matter induced nuclear recoil
rate due to the rotation of the Earth around the
Sun~\cite{Drukier:1986tm, Freese:1987wu}. The upgraded DAMA/LIBRA
detector confirmed~\cite{Bernabei:2008yi} the earlier result adding
many more statistics, and it has reached a significance of $8.9\sigma$
for the cumulative exposure~\cite{Bernabei:2010mq}.  In 2010, the
CoGeNT Collaboration reported an irreducible excess in the counting
rate~\cite{Aalseth:2010vx}, which may also be ascribed to a dark
matter signal. One year later, the same collaboration reported further
data analyses showing that the time series of their rate is actually
compatible with an annual modulation effect~\cite{Aalseth:2011wp}. In
CoGeNT data the evidence for the annual modulation is at the
$2.8\sigma$ level. In the summer of 2011, the CRESST Collaboration
also reported an excess of low energy events that are not consistent
with known backgrounds~\cite{Angloher:2011uu}. In particular, 67
counts were found in the dark matter acceptance region and the
estimated background from leakage of $e/\gamma$ events, neutrons,
$\alpha$ particles, and recoiling nuclei in $\alpha$ decays is not
sufficient to account for all the observed events. The CRESST
Collaboration rejected the background-only hypothesis at more than $4
\sigma$.  Of particular interest here, the DAMA (after including
the effect of channeling in the NaI crystal
scintillators~\cite{Petriello:2008jj}) and CoGeNT results appear to be
compatible with a relatively light dark matter particle, in the few
GeV to tens of GeV mass range, with a scattering cross section against
nucleons of about $7 \times 10^{-41}~{\rm
  cm}^2$~\cite{Fitzpatrick:2010em,Chang:2010yk,Hooper:2010uy,Buckley:2010ve}.
The central value favored by CRESST data points to somewhat larger
dark matter masses, but it is still compatible at the $1\sigma$ level
with the range determined by the other two experiments.

Very recently, CDMS II Collaboration reported three candidate events
with an expected background of 0.7 events~\cite{Agnese:2013rvf}.  If interpreted as a signal
of elastically scattering dark matter, the central value of the
likelihood analysis of the measured recoil energies favors a mass of
$8.6~{\rm GeV}$ and a scattering cross section on nucleons of 
\begin{equation}
\sigma_{w N}^{m_w \approx 10~{\rm GeV}} \approx 1.9
\times 10^{-41}~{\rm cm}^{2} \, . 
\label{CDMS}
\end{equation}
The 68\% confidence band is somewhat large and overlaps with previous signal
claims.

Alongside these ``signals'' stands the series of null results from the
XENON-100~\cite{Aprile:2011hi} and XENON-10~\cite{Angle:2011th}
experiments, which at present have the world's strongest exclusion
limit.  Some authors have pointed out that uncertainties in the
response of liquid xenon to low energy nuclear recoil may be
significant, particularly in the mass region of
interest~\cite{Collar:2010ht,Collar:2011wq}. In light of these
suspicions, a recent reanalysis of XENON data suggests candidates in
fact may have been observed~\cite{Hooper:2013cwa}. The data favor a
mass of 12~GeV, though the 90\% error contours extend from 7 to 30 GeV
with the cross section varying between $6\times 10^{-41}~{\rm cm}^2$
and $4 \times 10^{-45}~{\rm cm}^2$. Taken together, these
different arguments suggest that the existing data set is not
inconsistent with a dark matter candidate of about 10~GeV.

The $wN$ cross section for elastic scattering is given by
\begin{equation}
\sigma_{wN} = \frac{4}{\pi} \frac{m_w^2 m_N^2}{(m_w + m_N)^2} \ \frac{f_p^2 + f_n^2 }{2} \,,
\label{sigmawN}
\end{equation}
where $N \equiv \frac{1}{2} (n+p)$ is an isoscalar nucleon, in the
renormalization-group-improved parton
model~\cite{Ellis:2000ds,Beltran:2008xg}. The effective couplings to
protons $f_p$ and neutrons $f_n$ are given by \beq f_{p,n} = \sum_{q =
  u,d,s} \frac{G_q}{\sqrt{2}} f_{Tq}^{(p,n)} \frac{m_{p,n}}{m_q} +
\frac{2}{27} f_{TG}^{(p,n)} \sum_{q = c,b,t} \frac{G_q}{\sqrt{2}}
\frac{m_{p,n}}{m_q},
\label{effective-coup}
\eeq where $G_q$ is the $W$-WIMP's effective Fermi coupling for a
given quark species, \beq \mathscr{L} = \frac{G_{q}}{\sqrt{2}} \bar
\psi_- \psi_- \bar \psi_q \psi_q \,, \eeq with $\psi_q$ the SM quark
field of flavor $q$. The first term in (\ref{effective-coup}) reflects
scattering with light quarks, whereas the second term accounts for
interaction with gluons through a heavy quark loop. The terms
$f_{Tq}^{(p,n)}$ are proportional to the matrix element, $\la \bar q q \ra$, of quarks in a nucleon,
and are given by \beqa
f^{p}_{Tu} = 0.020 \pm 0.004, \quad \ f^{p}_{Td} = 0.026 \pm 0.005, \quad f^{p}_{Ts} = 0.118 \pm 0.062, \nonumber  \\
f^{n}_{Tu} = 0.014 \pm 0.003, \quad f^{n}_{Td} = 0.036 \pm 0.008,
\quad f^{n}_{Ts} = 0.118 \pm 0.062 \, .  \eeqa We also have
$f^{(p,n)}_{TG} = 1 - \sum_{u,d,s} f^{(p,n)}_{Tq}$, which is  $f^{p}_{TG} \approx 0.84$ and $f^{n}_{TG} \approx 0.83$~\cite{Ellis:2000ds}.

To establish the value of $G_{q}/m_q$ we look back at
(\ref{eq:fifty})  along with the SM Yukawa interaction term, which involves the mixing
of both scalar fields, $H$ and $h$. For interactions of $W$-WIMPs with SM quarks, the
relevant terms are \beq \mathscr{L} = \frac{m_q \cos \chi}{v_\phi} H \bar
\psi_q \psi_q - \frac{m_q \sin \chi}{v_\phi} h \bar \psi_q \psi_q +
\dots +\frac{f \sin \chi}{2} H \bar \psi_- \psi_- + \frac{f \cos
  \chi}{2} h \bar \psi_- \psi_-.  \eeq The scattering of a $w$
particle off a quark then gives \beqa \CM &=& i \frac{f m_q \sin \chi
  \cos \chi}{v_\phi} \bar u_q (p') u_q(p) \lpa \frac{1}{t-m_H^2} -
\frac{1}{t-m_h^2} \rpa \bar u(k') u(k)  \nonumber
\\
&\approx& i \frac{f m_q \eta_\chi v_r }{m_H^2 m_h^2} \bar u_q (p')
u_q(p) \bar u(k') u(k)  \nonumber
\\
&\approx& i \frac{m_q \eta_\chi \Delta m }{2 m_H^2 m_h^2} \bar u_q
(p') u_q(p) \bar u(k') u(k)  .  \eeqa This leads to the identification
of the effective coupling \beq  \frac{2 G_{q}}{\sqrt{2}} = \frac{m_q \eta_\chi
  \Delta m }{2 m_H^2 m_h^2} \Rightarrow \frac{G_{q}}{m_q} =
\frac{\eta_\chi \Delta m }{2\sqrt{2} \ m_H^2 m_h^2}.
\label{eq:effectiveCoup}
\eeq 
The insertion of  (\ref{eq:effectiveCoup}) and (\ref{effective-coup}) into
(\ref{sigmawN}) yields \beq
\sigma_{wN} \approx  3 \times 10^{-7} \,  \left[
  \frac{226.27 \ \eta_\chi \Delta m \ {\rm GeV}}{m_h^2} \right]^2~{\rm
  pb} \, .
\label{NOB}
\eeq
 
Combining (\ref{NOB}) with the signals/bounds on elastic scattering of dark matter particles on nucleons we obtain a constraining relation for $\eta_\chi \Delta m$. For $m_w = 10~{\rm GeV}$, we use the cross section reported by the CDMS Collaboration (\ref{CDMS}) to obtain 
\beq
\eta_\chi \Delta m = \frac{3.5 \times 10^{-2}}{\rm GeV} \   m_h^2    \, .
 \label{eq:DetAcdms}
\eeq 
For $m_w = 50~{\rm GeV}$, we adopt the 90\%~C.L. upper limit reported by
the XENON-100 Collaboration~\cite{Aprile:2011hi} to obtain 
\begin{equation}
   \eta_\chi \Delta m <  \frac{3.6 \times 10^{-4}}{\rm GeV} \   m_h^2
   \, .
 \label{eq:DetAxenon}
\end{equation}

\section{Constraints from Cosmological Observations}

The concordance model of cosmology predicts the evolution of a
spatially flat expanding Universe filled with dark energy,
dark matter, baryons, photons, and three
flavors of left-handed ({\it i.e.} one helicity state $\nu_L$)
neutrinos (along with their right-handed antineutrinos $\bar
\nu_R$). The
universal expansion rate, quantified by the Hubble parameter $H$, is
determined by the total energy density $\rho$,
\begin{equation}
  H^2 \equiv \frac{\dot a}{a} =  \frac{8 \, \pi \, G}{3}  \ \rho \,,
\label{Hubble}
\end{equation}
where $a$ is the expansion scale factor and $G$ is the gravitational
constant. In the relatively late, early Universe, the energy density is
dominated by radiation, that is by the contributions from massless
and/or extremely relativistic particles ({\it i.e.} $\rho \approx \rho_{\rm R}$).

The earliest observationally verified landmarks -- big bang
nucleosynthesis (BBN) and the cosmic microwave background (CMB)
decoupling epoch -- have become the de facto worldwide standard for
probing theoretical scenarios beyond the SM  containing new light
species. It is advantageous to normalize the extra contribution to the
SM energy density to that of an ``equivalent'' neutrino species. The
number of equivalent light neutrino species,
\begin{equation}
N_{\rm eff} = \frac{\rho_{\rm R} - \rho_\gamma}{\rho_{\nu_L}} \,,
\end{equation}
quantifies the total ``dark'' relativistic energy density (including
the three left-handed SM neutrinos) in units of the density of a
single Weyl neutrino 
\begin{equation}
\rho_{\nu_L} = \frac{7 \pi^2}{120} \ \left(\frac{4}{11} \right)^{4/3} T_\gamma^4, 
\end{equation}
where $\rho_\gamma$ is the energy density of photons (which by today
have redshifted to become the CMB photons at a temperature of about
$T_\gamma^{\rm today} \simeq 2.7~{\rm K}$)~\cite{Steigman:1977kc}.

\begin{figure}[ht]
\centering
\includegraphics[width=0.9\textwidth]{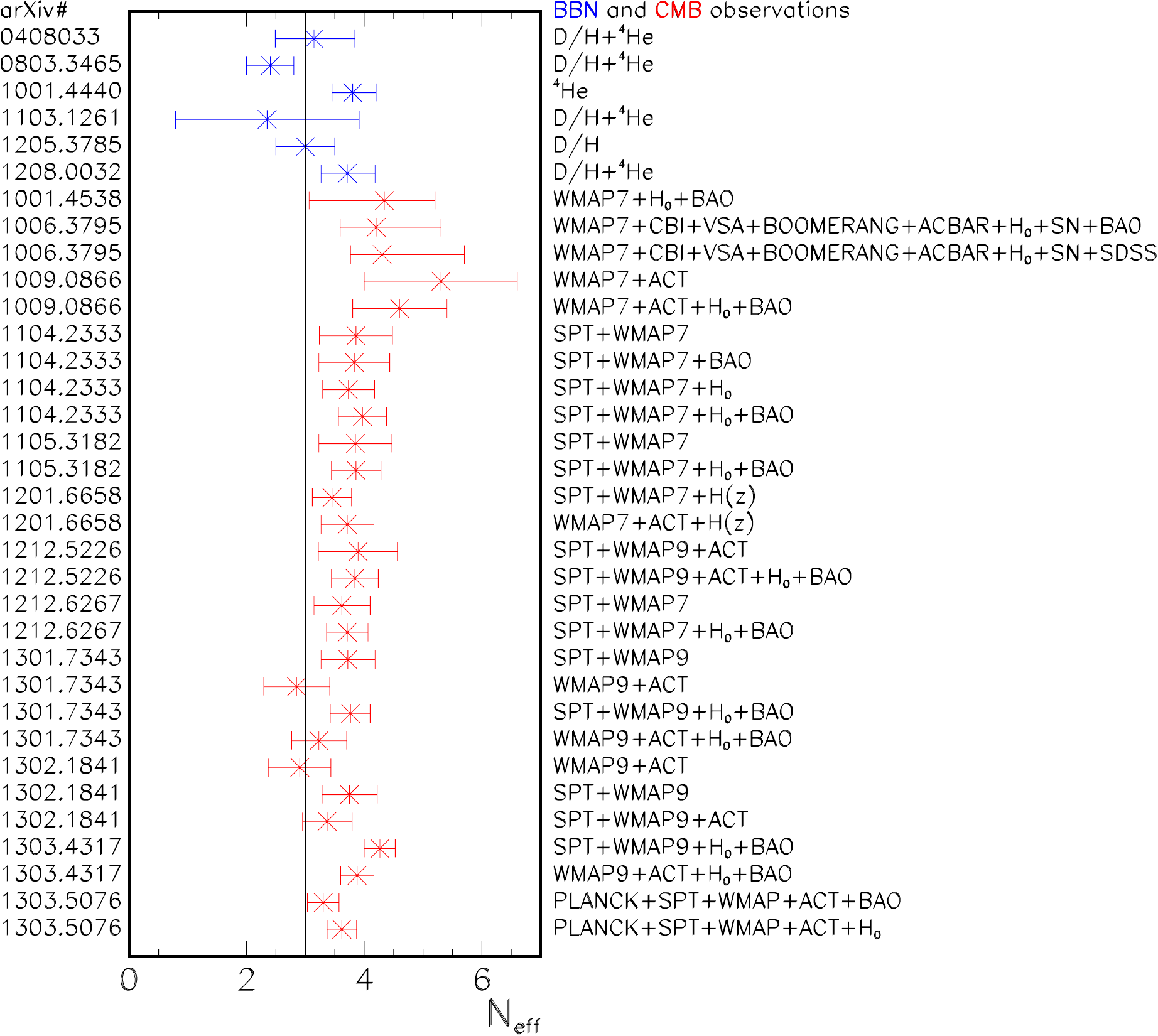}
 \caption{A selection of the most recent cosmological $N_{\rm eff}$
 measurements and the $1\sigma$ confidence intervals from various
 combinations of models and data sets. BBN findings~\cite{Cyburt:2004yc,Simha:2008zj,Izotov:2010ca,Mangano:2011ar,Pettini:2012ph,Steigman:2012ve} are shown in blue and those from the CMB epoch~\cite{Komatsu:2010fb,GonzalezGarcia:2010un,Dunkley:2010ge,Hou:2011ec,Keisler:2011aw,Moresco:2012jh,Hinshaw:2012fq,Hou:2012xq,DiValentino:2013mt,Calabrese:2013jyk,Benetti:2013wla,Ade:2013lta} in red.}
\label{fig:uno}
\end{figure}

 A selection of the most recent measurements of $N_{\rm eff}$ together
 with the $1\sigma$ confidence intervals from various combinations of
 models and data sets are shown in Fig.~\ref{fig:uno}. Altogether, the
 data hint at the presence of an excess $\Delta N$ above SM
 expectation of $N_{\rm eff} = 3.046$~\cite{Mangano:2005cc}.
 Arguably, one of the most intriguing results of the Planck spacecraft
 is that the best-fit Hubble constant has the value $h= 0.674 \pm
 0.012$~\cite{Ade:2013lta}.\footnote{We adopt the usual convention of
   writing the Hubble constant at the present day as $H_0 = 100 \
   h~{\rm km} \ {\rm s}^{-1} \ {\rm Mpc}^{-1}$.  There is some notational overlap with the differences of the massive states of the $r$ and $\phi$ fields; however, the context should make it clear what we are using.}  This result deviates
 by more than 2$\sigma$ from the value obtained with the Hubble Space
 Telescope, $h = 0.738 \pm 0.024$~\cite{Riess:2011yx}.  The impact of
 the new $h$ determination is particularly complex in the
 investigation of $N_{\rm eff}$. Combining CMB observations with data
 from baryon acoustic oscillations (BAO)~\cite{Percival:2009xn}, the
 Planck Collaboration reported $N_{\rm eff} = 3.30 \pm 0.27$. Adding the $H_0$
 measurement to the CMB data relieves the tension between the CMB data
 and $H_0$ at the expense of new neutrinolike physics (at around the
 $2.3 \sigma$ level), that is $N_{\rm eff} =3.62 \pm 0.25$.

 As noted in~\cite{Weinberg:2013kea}, the Goldstone boson $\alpha$ is a
 natural candidate for an imposter equivalent neutrino. The
 contribution of $\alpha$ to $N_{\rm eff}$ is $\Delta N =
 \rho_\alpha/\rho_{\nu_L}$.  Thus, taking into account the
 isentropic heating of the rest of the plasma between $T_{\alpha}^{\rm
   dec}$ and $T_{\nu_L}^{\rm dec}$ decoupling temperatures we obtain
\begin{equation}
\Delta N = \frac{4}{7} \left( \frac{g (T^{\rm dec}_{\nu_L})}{g (T^{\rm dec}_\alpha)} \right)^{4/3} \,,
\end{equation}
where $g(T)$ is the effective number of interacting (thermally
coupled) relativistic degrees of freedom at temperature $T$; for
example, $g(T_{\nu_L}^{\rm dec}) = 43/4$~\cite{Kolb:1990vq}.\footnote{
  If relativistic particles are present that have decoupled from the
  photons, it is necessary to distinguish between two kinds of $g$:
  $g_\rho$, which is associated with the total energy density, and
  $g_s$, which is associated with the total entropy density.  For our
  calculations we use $g = g_\rho = g_s$.}  For the particle content
of the SM, there is a maximum of $g(T_{\alpha}^{\rm dec}) =
427/4$ (with $T_{\alpha}^{\rm dec} > m_t$), which corresponds
to a minimum value of $\Delta N_\nu = 0.027$. 

We now turn to calculating the interaction rate for Goldstone bosons,
\beq
\Gamma  (T) = \sum_{\rm fermions} n_{\rm f}(T) \langle \sigma v \rangle \,,
\label{eq:twenty}
\eeq where 
\beq
n_{\rm f}(T) = \frac{ g_{\rm f} }{2 \pi^2} \int_0^\infty \frac{k^2}{e^{\beta \sqrt{k^2+m_{\rm f}^2}}+1} dk
\label{eq:twentyone}
\eeq is the number density of an interacting fermion of type f (with
mass $m_{\rm f}$), $\beta = (k_B T)^{-1}$, $g_{\rm f}$ is the number
of chiral states, and we average the cross section over the
statistical distribution for a given temperature. For $T \gg m_{\rm
  f}$, we obtain \beq n_{\rm f}(T) \approx g_{\rm f}
\frac{3 \zeta(3)}{4 \pi^2} \left( \frac{k_B T}{\hbar c} \right)^3.
\label{eq:twentytwo}
\eeq
This results in a simplification of (\ref{eq:twenty}) 
\beq
\Gamma (T) \approx   \frac{3 \zeta(3)}{4 \pi^2} \left( \frac{k_B T}{\hbar c} \right)^3 \sum_{\rm fermions} g_{\rm f} \langle \sigma v \rangle.
\label{eq:twentythree}
\eeq

Since the Goldstone boson only interacts with the SM fields via the
Higgs, we can have scatterings of the type $\alpha \psi \to \alpha \psi$,
with $\psi$ a generic SM fermion. The $\alpha$ scattering off fermions is 
described by SM Yukawa interaction terms that can be written  as
\beqa Y_{\rm f} \phi \bar{\psi} \psi & \rightarrow & Y_{\rm f} v_\phi \bar{\psi} \psi +
Y_{\rm f} \phi' \bar{\psi} \psi , \nonumber
\\
&=& m_{\rm f} \bar{\psi} \psi + \frac{m_{\rm f}}{v_\phi}  H \bar{\psi}
\psi  \, \cos \chi - \frac{m_{\rm f}}{v_\phi} h \bar{\psi}
\psi  \, \sin \chi \,,
\label{eq:fourteen}
\eeqa where $Y_{\rm f}$ is the Yukawa coupling  of the fermion in question. 

We proceed to calculate the scattering cross section.  The invariant
amplitude follows from the Feynman rules
\begin{equation}
i \CM = \frac{2  m_f \sin \chi \cos \chi}{v_r v_\phi} (k\cdot k') \frac{i}{t
  - m_H^2} \bar{u}(p') u(p) -\frac{2 m_f \cos \chi \sin \chi}{v_r v_\phi} (k\cdot k') \frac{i}{t
  - m_h^2} \bar{u}(p') u(p)  . 
\end{equation}
The momenta  of   incoming
and outgoing (outgoing primed) particles are defined by
\beqa
p^\mu &=& \left(p, p \sin \varphi, 0, -p \cos \varphi \right) \nonumber
\\
k^\mu &=& \left(k, 0, 0, k \right) \nonumber
\\
k'^\mu &=& \left(k', k' \sin \vartheta, 0, k' \cos\vartheta \right) \nonumber
\\
p'^\mu &=& \left(p', -p' \sin \vartheta', 0, - p' \cos \vartheta' \right) \,,
\label{eq:twentysix}
\eeqa 
with  $t = p'-p$.  To obtain the (unpolarized) cross section, we have
to take the square of the modulus of ${\cal M}$ and then carry out the
spin and color (if appropriate) sums
\begin{equation}
\frac{1}{2} \sum_{\rm spins,\;  colors} |\CM|^2 = 8 N_c \lpa \frac{m_f \sin \chi \cos \chi}{v_r v_\phi} \rpa^2 \lpa \frac{m_H^2-m_h^2}{(t-m_H^2)(t-m_h^2)} \rpa^2 (k\cdot k')^2 (p\cdot p' + m_f^2) ,
\label{eq:seventeenW}
\end{equation}
where $N_c =3$ for quarks and $N_c =1$ for leptons. 
The cross section in the center-of-mass (c.m.) frame in the highly relativistic approximation is given by
\beq
\frac{d\sigma}{d \Omega} \approx \frac{N_c}{8 \pi^2 s}\lpa \frac{m_f \eta_\chi}{(t-m_H^2)(t-m_h^2)} \rpa^2 (k\cdot k')^2 (p\cdot p') \,,
\label{eq:seventeen}
\eeq where $s = (k + p)^2 \approx 4 k^2$ and finally $\eta_\chi \ll
1$. To make progress on this problem we take the effective coupling
form \beq \sigma (s) \approx \frac{N_c}{64 \pi} \lpa \frac{m_f
  \eta_\chi}{m_H^2 m_h^2} \rpa^2 s^2 \, . \eeq

Nonequilibrium thermal physics tells us that the way to do thermal
averaging within Boltzmann's  approximation is
\begin{eqnarray}
\la \sigma v \ra & = & \int d\Pi_{p'} d\Pi_{k'} d\Pi_{k} d\Pi_{p}
|\CM(k+p\rightarrow k' + p')|^2 f_{\rm f}(p, T) f_\alpha(k, T) \nonumber \\
& \times & (2\pi)^4 \delta^{(4)} (p+k-p'-k'),
\label{assman}
\end{eqnarray} 
with $d\Pi_{p} = d^3 p'/[(2\pi)^3 2 E_{p'}]$ and likewise for the
other parameters.  Here, $f_{\rm f}$ and $f_\alpha$ are Fermi and Bose
equilibrium normalized distributions, corresponding to the ${\rm f}$
fermion and $\alpha$ boson, respectively.  The expression from
nonequilibrium thermal physics [Eq.~(\ref{assman})] is approximated
by \beq \la \sigma v \ra \approx \int \frac{d^3 k}{(2\pi)^3}\frac{d^3
  p}{(2\pi)^3} f_{\rm f}(p, T) f_\alpha(k, T) \ v_M \ \sigma(s) \, ,
\eeq where $v_M \approx k \cdot p/(pk) = 2 (1+\cos \varphi)$ is the
M\"oller velocity in the ultrarelativistic
limit~\cite{Gondolo:1990dk,Weiler:2013hh} and $s = 2 k p (1+\cos
\varphi)$ is the c.m. energy of two interacting particles with initial
momenta not necessarily collinear.  The velocity average cross section
then is found to be \beqa \la \sigma v \ra &\approx& \frac{1}{8 \pi^4}
\int_0^\infty p^2 dp \int_0^\infty k^2 dk \int_0^\pi \sin \varphi d
\varphi \ f_{\rm f}(p,T) f_\alpha(k,T) \nonumber \\ & \times & 2 (1+
\cos\varphi) \ \, \sigma_{\rm c.m.} [2 k p (1+\cos \varphi)],
\nonumber
\\
&=& N_c \frac{15 \zeta^2(5)}{\pi \zeta^2(3)} \lpa \frac{m_f
  \eta_\chi}{m_H^2 m_h^2} \rpa^2 (k_B T)^4, \nonumber
\\
&\approx& 3.55 N_c \lpa \frac{m_f \eta_\chi}{m_H^2 m_h^2} \rpa^2 (k_B
T)^4.  \eeqa Putting this all together, we obtain \beq \Gamma (T)
\approx 0.32 \lpa \frac{\eta_\chi}{m_H^2 m_h^2} \rpa^2 (k_B T)^7
\sum_{\rm fermions} g_{\rm f} \, N_{\rm c} \, m_{\rm f}^2.  \eeq

Now, since we can approximate the energy density (at high temperatures) by
including only particles species $i$ with $T \gg m_i$, it follows that
\begin{equation}
\rho_{\rm R} = \left(\sum_{\rm bosons} g_{\rm b} + \frac{7}{8} \sum_{\rm fermions} g_{\rm f} \right) \frac{\pi^2}{30} (k_BT)^4 = \frac{\pi^2}{30} g(T) (k_BT)^4
\end{equation}
and therefore the Hubble parameter (\ref{Hubble}) becomes
\begin{equation}
  H(T)  \simeq \frac{1.66}{M_{\rm Pl}}
  \sqrt{g(T)} \ (k_B T)^2  \, ,
\end{equation}
where $g_{\rm b  (f)}$ is the number of degrees of freedom of each boson (fermion) and the sum runs over all boson and fermion states with $T
\gg m_i$. The factor of 7/8 is due to the difference between the Fermi
and Bose integrals.

The Goldstone boson decouples from the plasma when its mean free path
becomes greater than the Hubble radius at that time
\begin{equation}
\Gamma (T_\alpha^{\rm dec}) =  H(T_\alpha^{\rm dec}) \, .
\end{equation}
The most interesting thermodynamics originates if $\alpha$ goes out of
thermal equilibrium while $T$ is still above the mass of the
muons but below the mass of all other particles of the SM, a time when
neutrinos are still in thermal equilibrium. For instance, with $\eta_\chi = 0.005$ and $m_h \approx 500~{\rm MeV}$ we obtain~\cite{Weinberg:2013kea}
\begin{equation}
\Delta N
= (4/7) (43/57)^{4/3} = 0.39 \, .
\end{equation}
This corresponds to a number of equivalent light neutrino species that
is consistent at the $1\sigma$ level with both the estimate of
$N_{\rm eff}$ using Planck + BAO data as well as the estimate using Planck +
$H_0$ data.

However, of particular interest here is the case where the mass of the
Goldstone boson companion field is $m_h \approx 98~{\rm GeV}$ and
$\eta_\chi = 0.0003$. For such set of parameters, $\alpha$ decouples
when 
\beq 0.32 \lpa \frac{\eta_\chi}{m_H^2 m_h^2} \rpa^2 \lpa k_B T
\rpa^7  12 \ m_b^2 = \frac{1.66}{M_{\rm
    Pl}}\sqrt{86.25}\lpa k_B T \rpa^2,
\label{eq:Nfixer}
\eeq 
where we have approximated $\sum_{\rm fermions} N_c \ g_{\rm f} 
\ m_{\rm f}^2 \approx 12 m_b^2$.  This gives $T \approx 5~{\rm GeV}$, and so the
$\alpha$ contribution to $N_{\rm eff}$ is found to be \beq \Delta N
\approx 0.036. \eeq The corresponding value of $N_{\rm eff}$ is within
the $1\sigma$ interval of the value reported by the Planck
Collaboration using Planck + BAO data but far out from the value
derived using Planck + $H_0$ data. Should future data point toward
the Planck + $H_0$ value, one should find a different origin to
explain the extra relativistic degrees of freedom (if $m_h \approx
98~{\rm GeV}$). One interesting possibility is to include the
right-handed partners of the three left-handed, SM neutrinos. It was
shown
elsewhere~\cite{Anchordoqui:2011nh,Anchordoqui:2012wt,SolagurenBeascoa:2012cz,Anchordoqui:2012qu,Anchordoqui:2013wwa}
that milliweak interactions of these Dirac states (through their
coupling to a TeV-scale $Z'$ gauge boson) may allow the $\nu_R$'s to
decouple during the course of the quark-hadron crossover transition,
just so that they are partially reheated compared to the $\nu_L$'s.
Remarkably, the required mass for the $Z'$ gauge boson is within the
range of discovery of LHC.

\section{Fitting Fermi data and the Observed Dark Matter Density}

Next, in line with our stated plan, we use \emph{Fermi} data and the
observed relic density to determine the free parameters of the
model. To this end we first calculate the annihilation rate into SM
fermions and Goldstone bosons.

\subsection{$\bm{W}$-WIMP Annihilation into SM Fermions}

The $W$-WIMP can annihilate into SM fermions  via $\bar{\psi}_- \psi_-
\rightarrow \phi^*/r^* \rightarrow \bar{\psi}{\psi}$, with an $s$-channel
Higgs or $h$ mediator. The matrix element of this process is
given by \begin{equation} i \CM = i f \sin \chi \, \cos\chi \
  \bar{v}(p') u(p) \left(\frac{i }{s - m_H^2} -
    \frac{i}{s-m_h^2}\right) \frac{i m_{\rm f}}{v_\phi} \bar{u}(k')
  v(k) \, . \end{equation} The minus sign in the second propagator
is necessary because the $r$ couples with a negative sign to fermions compared
to the Higgs; see~(\ref{eq:fourteen}). The spin-averaged invariant
amplitude reads \begin{equation} \frac{1}{4}\sum|\CM|^2 = 
  N_c \left( \frac{f m_{\rm f} \sin \chi \, \cos \chi}{v_\phi}
  \right)^2\frac{4 \, \left( m_h^2 - m_H^2 \right)^2
     (p \cdot p'-m_w^2)(k \cdot k' - m_{\rm f}^2)}{(s-m_h^2)^2(s-m_H^2)^2} \, .
\end{equation}
Now, let us calculate the cross section for ${\rm f} \bar {\rm f}$-pair production
\beqa
d\sigma  &=& \frac{1}{8 E_p |p| } |\CM|^2 \frac{d^3 k}{(2\pi)^3 2 E_k} \frac{d^3 k'}{(2\pi)^3 2 E_{k'}} (2\pi)^4 \delta^{(3)}(\textbf{k}'+\textbf{k}) \nonumber \\
& \times & \delta(2 E_p - E_k - E_{k'}), 
\end{eqnarray}
and so 
\begin{eqnarray}
  \sigma &=& \frac{|\CM|^2}{64 \pi } \frac{|k'|}{|p| E_p^2} \nonumber
  \\
  &=&  \frac{N_c}{16 \pi}  \, \left( \frac{f m_{\rm f} \sin \chi \ \cos \chi}{v_\phi} \right)^2\frac{|k'|}{|p|} \frac{ \left( m_h^2 - m_H^2\right)^2}{(s-m_h^2)^2(s-m_H^2)^2}  \frac{\lpa p \cdot p' -m_w^2 \rpa \lpa k \cdot k' - m_{\rm f}^2 \rpa}{ E_p^2}  \nonumber
  \\
  &\approx&   \frac{N_c}{16 \pi} \left( \frac{\eta_\chi m_{\rm f}  \Delta m }{2(s-m_h^2)(s-m_H^2)} \right)^2 \sqrt{\frac{|s-4m_{\rm f}^2|}{|s-4m_w^2|}} \frac{(s-4m_w^2)(s-4m_{\rm f}^2)}{s}\, .
\label{CDMx}
\end{eqnarray}  
In this case the out
state does not consist of identical particles. For phenomenological purposes, the $h$ pole needs to be softened to a
Breit--Wigner form by obtaining and utilizing the correct total widths
$\Gamma_h$ of the resonance.  This is accomplished by modification of the $s$-channel propagator for $h$ via
\beq
\frac{i}{s -m_h^2} \rightarrow \frac{i}{s- m_h^2- i m_h\Gamma_h} \, .
\eeq
After this is done, the contribution of the ${\rm f} \bar {\rm f}$ channel is as follows:  
\beqa \sigma &=& \frac{N_c}{16 \pi} \left( \frac{\eta_\chi
    m_{\rm f} \Delta m }{2 (m_H^2-m_h^2) (s-m_H^2)} \right)^2 \frac{
  \left( m_H^2 - m_h^2\right)^2 + m_h^2 \Gamma_h^2}{(s-m_h^2)^2 +
  m_h^2 \Gamma_h^2}  \sqrt{\frac{|s-4m_{\rm f}^2|}{|s-4m_w^2|}}
\nonumber \\
& \times & \frac{(s-4m_w^2)(s-4m_{\rm f}^2)}{s} \ , \nonumber \\
&\approx& \frac{N_c}{16 \pi} \left( \frac{\eta_\chi
    m_{\rm f} \Delta m }{2  (s-m_H^2)} \right)^2 \frac{
  1 }{(s-m_h^2)^2 +
  m_h^2 \Gamma_h^2}  \sqrt{\frac{|s-4m_{\rm f}^2|}{|s-4m_w^2|}}
\frac{(s-4m_w^2)(s-4m_{\rm f}^2)}{s}  .
\eeqa 
For $\Delta m > m_H/2$, the decay channels of the $h$ field are $h \rightarrow {\rm f} \bar {\rm f}$, $h \rightarrow w \bar w$, and $h
\rightarrow 2 \alpha'$. The corresponding decay widths are given by
\beqa \Gamma_{h\rightarrow {\rm f} \bar {\rm f} } &=& \sum_{\rm
  fermions} \frac{N_c}{8 \pi m_h^2} \lpa \frac{m_{\rm f} \sin \chi}{v_\phi} \rpa^2 (m_h^2-4 m_{\rm f}^2)^{3/2} \nonumber
\\
&\approx& \sum_{\rm fermions}  \frac{N_c}{8 \pi m_h^2} \lpa \frac{m_{\rm f} \eta_\chi v_r}{m_H^2-m_h^2} \rpa^2 (m_h^2-4 m_{\rm f}^2)^{3/2} \nonumber
\\
&\approx& \sum_{\rm fermions}  \frac{N_c}{8 \pi m_h^2 f^2} \lpa \frac{m_{\rm f} \eta_\chi \Delta m}{2(m_H^2-m_h^2)} \rpa^2 (m_h^2-4 m_{\rm f}^2)^{3/2} \nonumber
\\
&\approx& \frac{3}{8 \pi m_h^2 f^2} \lpa \frac{m_{\rm b} \eta_\chi \Delta m}{2(m_H^2-m_h^2)} \rpa^2 (m_h^2-4 m_{\rm f}^2)^{3/2}
\eeqa (in the last line we have taken
$m_b < m_w < m_t$), \beqa \Gamma_{h\rightarrow w \bar w} &=& \frac{2
  \, ( m_h^2- 4 m_w^2) \, f^2 \cos^2 \chi }{32 \pi m_h^2}
\sqrt{m_h^2-4m_w^2} \nonumber
\\
&\approx& \frac{f^2}{16 \pi m_h^2} ( m_h^2- 4 m_w^2)^{3/2} \eeqa
(inclusion of this channel requires $2 m_w < m_h$), and \beqa
\Gamma_{h\rightarrow 2 \alpha'} &=& \frac{1}{32 \pi}\lpa \frac{\cos
  \chi}{v_r} \rpa^2m_h^3 \nonumber
\\
&\approx& \frac{f^2 }{8 \pi \Delta m^2} m_h^3 \, .  \eeqa The dominant 
terms of the total decay width come from the hidden sector. Hence, in what
follows we  neglect  terms accounting for $h$ decay into the visible
sector and consider $m_h < 2 m_w$  (so that the decay $h \rightarrow w 
\bar w $ is closed). Under these assumptions  the decay width takes a particularly simple form
\beq \Gamma_h = \frac{f^2 }{8 \pi \Delta m^2} m_h^3  \, .
\label{finalform}
\eeq

Next, we compute the averaged cross section for thermal interactions.
In the cosmic comoving frame (the frame where the gas is assumed to be
at rest as a whole) we have \beq \la \sigma v \ra = \frac{\int d^3 p
  d^3 p' f_w(p, T) f_w(p', T) \, \sigma v_M}{\int d^3 p d^3 p' f_w(p,
  T) f_w(p', T) }, 
\label{eq:ThermAve}
\eeq where $\bm{p}$ and $\bm{p}'$ are the
three-momenta of the colliding particles, whose equilibrium
distribution function at temperature $T$ is Maxwell--Boltzmann, \beq
f_w(p,T) \approx e^{- \beta \sqrt{p^2 + m_w^2}} \, , \eeq with $p =
|\bm{p}|$ and $p' = |\bm{p}'|$. The Maxwell--Boltzmann distribution
remains a good approximation
provided $3 \, m_w \, \beta \agt 1 $. The M\"oller velocity can be expressed as
\beq v_M = \frac{1}{E E'} \sqrt{(p\cdot p')^2 - m_w^4} = \frac{1}{2 E
  E'} \sqrt{s (s-4m_w^2)} \, ,
\label{nonrelmoller}
\eeq where $E$ and $E'$ are the energies of the scattering
particles. Note that in the c.m. frame the velocity of the colliding
$W$-WIMPs is half the M\"oller velocity, $v = \sqrt{1-4m_w^2/s} =
v_M/2$.

For $s \gg m_{\rm f}$, from (\ref{CDMx}) and (\ref{nonrelmoller}) we
obtain
\begin{equation}
\sigma v_M = \frac{N_c}{8 \pi} \left( \frac{\eta_\chi m_{\rm f}  \Delta m }{2(s-m_h^2)(s-m_H^2)} \right)^2 \,  (s - 4 m_w^2) \, .
\end{equation}
We evaluate (\ref{eq:ThermAve}) by expanding $\sigma v_M$ around \beq
s = 4 E^2 = \frac{4 m_w^2}{1-v^2} \approx 4m_w^2 (1 + v^2 + \dots) \
 \eeq to obtain a series solution in powers of $v$ of which the
leading order term is \beq \la \sigma v \ra \approx \frac{N_c}{2\pi}
\lpa \frac{ \eta_\chi m_{\rm f} m_w \Delta
  m}{2(4m_w^2-m_h^2)(4m_w^2-m_H^2)} \rpa^2 \la v^2 \ra \,,\eeq where
$\la v^2 \ra$ is the $W$-WIMP thermally averaged velocity.  

All in all, the total average annihilation cross section
into SM particles (labelled by subindex $i$) is given by \beq
\sum_{\rm fermions} \la \sigma_i v \ra \approx \frac{ 3 }{ 2 \pi }
\left( \frac{\eta_\chi m_b m_w \Delta m }{2 (4 m_w^2-m_H^2)} \right)^2
\, \frac{\la v^2 \ra}{(4 m_w^2-m_h^2)^2 + m_h^2 \Gamma_h^2} \, ,
\label{eq:TEWII}
\eeq 
where we have assumed that the overwhelming contribution into $b \bar b$ dominates the process.

Provided the theory is not strongly coupled, (\ref{eq:TEWII}) is
generally a good approximation for relativistic particles, but for low
velocities and in the presence of a long-range force (classically,
when the potential energy due to the long-range force is comparable to
the particles' kinetic energy), the perturbative approach breaks
down. In the nonrelativistic limit, the question of how the long-range
potential modifies the cross section for short-range interactions can
be formulated as a scattering problem in quantum mechanics, with
significant modifications to the cross sections occurring when the
particle wave functions are no longer well approximated by plane waves
(so the Born expansion is not well behaved). The deformation of the
wave functions due to a Coulomb potential was calculated by
Sommerfeld~\cite{Sommerfeld:1931}, yielding a $\sim 1/v$ enhancement
to the cross section for short-range interactions (where the
long-range behavior due to the potential can be factorized from the
relevant short-range behavior). Along these lines, for low-velocity
($v \sim 10^{-3}$) $W$-WIMPs in our Galactic halo, we expect
interactions with the $H$ and $h$ fields to enlarge the cross section,
as the attractive Yukawa potential, \beq V_w(r) = -\frac{f^2
  \cos^2(\chi)}{4\pi} \frac{e^{-m_h r}}{r} - \frac{f^2
  \sin^2(\chi)}{4\pi} \frac{e^{-m_H r}}{r} \approx -\frac{f^2}{4\pi}
\frac{e^{-m_h r}}{r} \simeq -\frac{f^2}{4\pi} \frac{1}{r} \ ,  \eeq
causes passing $W$-WIMPS to be drawn toward each
other~\cite{ArkaniHamed:2008qn}.

For $p$-wave scattering, $ \la v^2
\ra \rightarrow \la S (v) v^2 \ra ,$ where
 \begin{equation}
S (v)  \approx   \frac{\pi \tilde{\alpha}}{v}\frac{1}{1-e^{- \pi
    \tilde{\alpha}/v}} \lpa 1+ \frac{\pi^2 \tilde{\alpha}^2}{4 v^2}
\rpa \ ,
 \label{eq:SomS1}
 \end{equation}
 is the Sommerfeld enhancement factor in the Coloumb approximation,
 with $\tilde \alpha = f^2/(4\pi)$~\cite{Cassel:2009wt}.
 Following~\cite{Feng:2010zp} we compute the thermally averaged
 Sommerfeld enhancement factor by approximating  $\lpa 1-e^{- \pi
   \tilde{\alpha}/v} \rpa^{-1}$  with $\tilde{\alpha} \ll 1$ \beq
 \la S (v) v^2 \ra \approx 6 x^{-1} + 4 \sqrt{\pi} \tilde{\alpha}
 x^{-1/2} + \frac{4 \pi^2 \tilde{\alpha}^2}{3} + \pi^{5/2}
 \tilde{\alpha}^3 x^{1/2} +\frac{\pi^4 \tilde{\alpha}^4}{6} x, \eeq
 where $x = m_w/T$. For interactions in the Galactic halo (G.h.), we have
 $\la v^2 \ra \sim 10^{-6}$, and therefore the thermally average
 annihilation cross section into $b \bar b$ becomes \beqa \la \sigma_b
 v \ra & \approx & \frac{ 3 }{ 2 \pi } \left( \frac{\eta_\chi m_b m_w
     \Delta m }{2 (4 m_w^2-m_H^2)} \right)^2 \, \frac{1}{(4
   m_w^2-m_h^2)^2 + m_h^2 \Gamma_h^2} \nonumber
 \\
 &\times& \frac{1}{4} \lpa6 x_{\rm G.h.}^{-1} + 4 \sqrt{\pi}
 \tilde{\alpha} x_{\rm G.h.}^{-1/2} + \frac{4 \pi^2
   \tilde{\alpha}^2}{3} + \pi^{5/2} \tilde{\alpha}^3 x_{\rm
   G.h.}^{1/2} +\frac{\pi^4 \tilde{\alpha}^4}{6} x_{\rm G.h.} \rpa \,,
 \eeqa with $x_{\rm G.h.}  \approx 3 \times 10^{6}$.

\subsection{$\bm{W}$-WIMP Annihilation into Pairs of Goldstone Bosons}

In addition to the annihilation into SM fermions we must consider the  $w \bar w
\rightarrow 2 \alpha'$ annihilation channel. The invariant amplitude for this process
is given by \beq i \CM = \frac{2 i f}{v_r} \bar{v}(p) u(p') \left(
  \frac{\sin^2 \chi}{s-m_H^2} - \frac{\cos^2 \chi}{s-m_h^2} \right) k
\cdot k'.  \eeq We then average over the in state spins to obtain
\beqa \frac{1}{4} \sum_{s,s'} |\CM|^2 &=& \frac{ f^2 s^2
  [(s-m_h^2)\sin^2 \chi - (s-m_H^2)\cos^2 \chi]^2}{2 v_r^2
  (s-m_h^2)^2(s-m_H^2)^2} (s - 4m_w^2) . \nonumber \eeqa The general
expression for the cross section reads \beq \sigma  =
\frac{1}{16 \pi \sqrt{s} \sqrt{|s-4m_w^2|} } \frac{f^4 s^2
  [(s-m_h^2)\sin^2 \chi - (s-m_H^2)\cos^2 \chi]^2}{\Delta m^2
  (s-m_h^2)^2(s-m_H^2)^2} (s - 4m_w^2) \, .  \eeq Using the small
angle approximation, {\it i.e.} $\cos \chi \approx 1$, we obtain \beq
\sigma  \approx \frac{f^2 s^2 \sqrt{|s-4m_w^2|}}{16 \pi
  \sqrt{s} (s-m_h^2)^2 } \lpa \frac{f^2}{\Delta m^2} +
\frac{(m_h^2+m_H^2-2 s)}{ 2 (s-m_H^2)^2}\frac{\eta_\chi^2
  v_\phi^2}{(m_H^2-m_h^2)^2} \rpa \ .  \eeq Taking a thermal average
gives \beq \la \sigma_{\alpha'} v \ra  \approx  \frac{2 f^4 m_w^4}{\pi \Delta m^2 [(m_h^2-4 m_w^2)^2 + m_h^2 \Gamma_h^2]} \la v^2 \ra \, .
\label{eq:siduerme}
\eeq
If the $W$-WIMPs are highly nonrelativistic we have to correct (\ref{eq:siduerme}) to account for the Sommerfeld enhancement, 
\beq \la \sigma_{\alpha'} v \ra  \approx  \frac{2 f^4 m_w^4}{
\pi \Delta m^2 [(m_h^2-4 m_w^2)^2 + m_h^2 \Gamma_h^2]} \la  S(v) v^2 \ra \, .
\eeq

\subsection{$\bm{W}$-WIMP Parameter Fits}

\begin{figure}[tbph]
\includegraphics[width=0.9\textwidth]{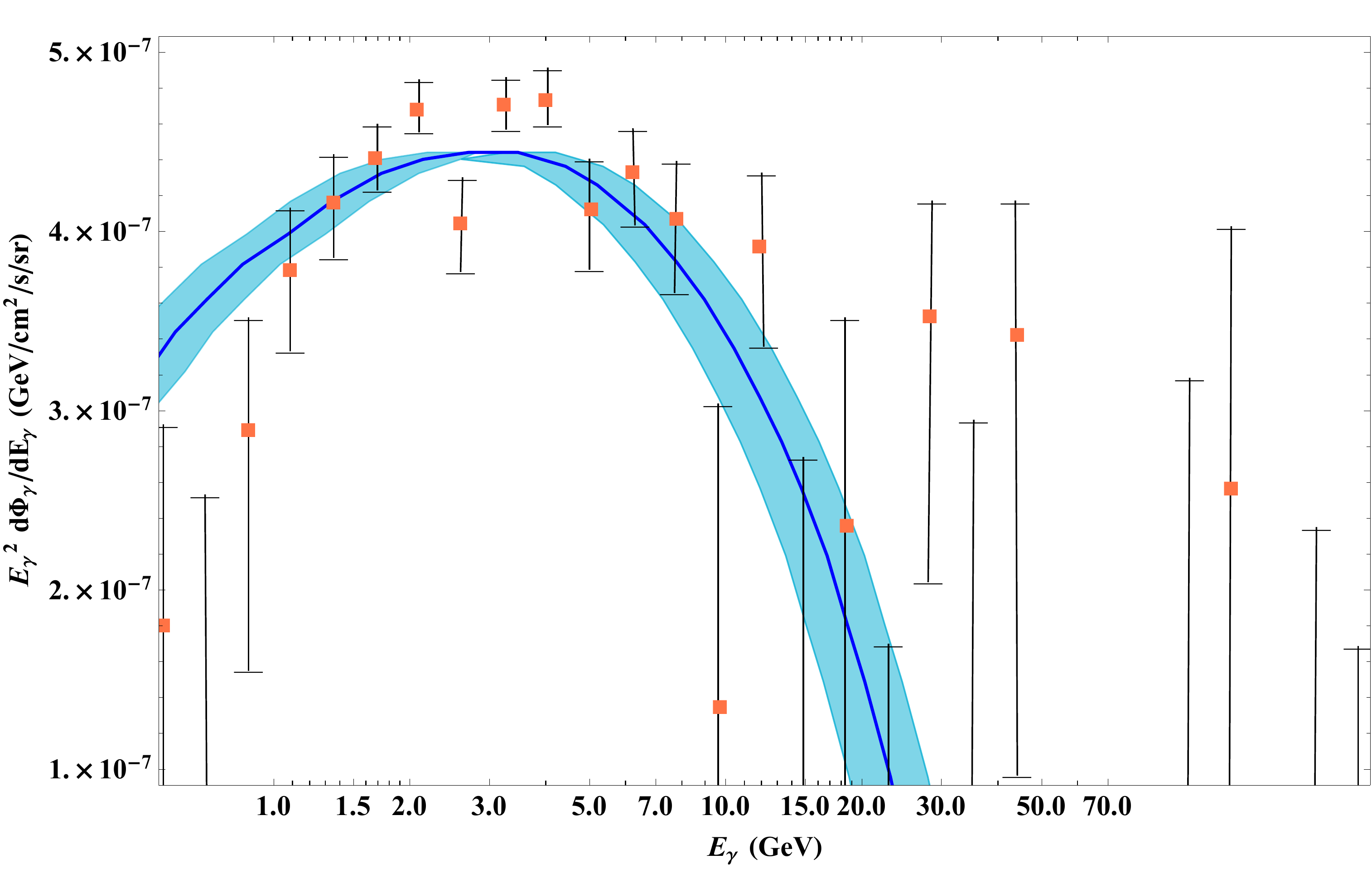}
\caption{Comparisons of the observed $\gamma$-ray spectrum of the
  low-latitude ($|b| = 10^\circ - 20^\circ$) emission, after
  subtracting the contribution from inverse Compton scattering to that
  predicted from 50~GeV $W$-WIMPs annihilating to $b \bar b$. We have adopted a generalized NFW profile with an inner slope of $\gamma = 1.2$, and normalized the signal to a local density of $0.4~{\rm GeV}/{\rm cm}^3$ and an annihilation cross section of $\langle \sigma_b v \rangle = 8 \times 10^{-27}~{\rm cm}^3/{\rm s}$. The band shows the variation in the mass range $45~{\rm GeV} < m_w < 55~{\rm GeV}$ for the same normalization. Adapted from Fig.~14 of Ref.~\cite{Hooper:2013rwa}.}
\label{fig:dos}
\end{figure} 

The total flux of $\gamma$-rays per solid angle from $W$-WIMP annihilation
into  SM particles (labelled by subindex $i$) is given by
\begin{equation}
\frac{d\Phi_\gamma}{dE_\gamma} = \sum_{\rm fermions} \frac{\langle
  \sigma_i v
  \rangle}{2} \ \frac{{\cal
  J}_{\Delta \Omega}}{J_0} \, \frac{1}{\Delta \Omega_{\rm obs} m_w^2} \,
\left. \frac{dN_\gamma}{dE_\gamma} \right|_i \,,
\end{equation}
where ${\cal J}/J_0$ is the normalized integral of mass density
squared of the dark matter in the line of sight, $dN_\gamma/dE_\gamma$ is the
$\gamma$-ray spectrum per annihilation into particle species $i$,
$\Delta \Omega_{\rm obs}$ is the observational solid angle in
steradians, and the sum runs over all possible annihilation
channels. It is noteworthy that $d\Phi_\gamma/dE_\gamma$ is the total photon
number flux per unit energy per unit steradian for a full sky
observation and, when compared to the total photon count of the
Fermi-LAT observation with $|b| > 10^\circ$, must be scaled to the field of
view of that observation, $\Delta \Omega_{\rm obs} = 10.4~{\rm sr}$.

{}From (\ref{eq:TEWII}) we see that, for $10~{\rm GeV} \alt m_w \alt
50~{\rm GeV}$, the dominant annihilation channel is $b \bar
b$. Annihilation into $c \bar c$ and $\tau^+ \tau^-$ is suppressed by
about 1 order of magnitude. Hereafter, we make the case for a $w$ with
a mass of about $50~{\rm GeV}$, which annihilates into $b \bar b$.
The photon flux expected from the \emph{Fermi} bubbles is shown in
Fig.~\ref{fig:dos}. Comparing (\ref{fermibubble2}) and
(\ref{eq:TEWII}), we obtain \beqa \la \sigma_b v \ra & \approx & \frac{
  3 }{ 2 \pi } \left( \frac{\eta_\chi m_b m_w \Delta m }{2 (4
    m_w^2-m_H^2)} \right)^2 \, \frac{1}{(4 m_w^2-m_h^2)^2 + m_h^2
  \Gamma_h^2} \, \frac{1}{4} \lpa6 x_{\rm G.h.}^{-1} + 4 \sqrt{\pi}
\tilde{\alpha} x_{\rm G.h.}^{-1/2} + \frac{4 \pi^2
  \tilde{\alpha}^2}{3}
\right. \nonumber \\
& + & \pi^{5/2} \tilde{\alpha}^3 x_{\rm G.h.}^{1/2} +
\left. \frac{\pi^4 \tilde{\alpha}^4}{6} x_{\rm G.h.} \right) = 6.7
\times 10^{-10} \rm GeV^{-2} \, .
 \label{eq:FermiBub}
\eeqa
 
To be produced thermally
in the early Universe in an abundance equal to the measured dark
matter density,  $\Omega_{\rm DM} h^2 = 0.1120 \pm 0.0056$~\cite{Beringer:1900zz}, the 50 GeV $w$ particle must have an annihilation 
cross section of
\beq
\sum_{\rm all \, species} \langle \sigma_i v \rangle \sim 3
\times 10^{-26}~{\rm cm}^3/{\rm s} = 2.5 \times 10^{-9}~{\rm GeV}^{-2} , 
\label{gary}
\eeq when thermally averaged over the process of
freeze-out, $x_{\rm f.o.} \sim 20$~\cite{Scherrer:1985zt,Steigman:2012nb}. It is noteworthy that for $\tilde \alpha \alt 0.01$ the effect of the Sommerfeld enhancement on the final relic particle abundance is negligible~\cite{Dent:2009bv,Chen:2013bi}. Herein we will work on the range of the coupling $\tilde \alpha$ over which Sommerfeld annihilation can be neglected in the calculation of relic densities.

Because  {\it a priori} we do not know whether $\langle
\sigma_{\alpha'} v\rangle$ or $\langle \sigma_b v\rangle$ dominates
the total annihilation cross section at freeze-out, we combine
(\ref{eq:TEWII}) and (\ref{eq:siduerme}) evaluated at $v(x_{\rm
  f.o.})$ together with (\ref{gary}) to obtain \beq \left[ \frac{2 f^4
    m_w^4}{ \pi \Delta m^2} +\frac{ 3 }{ 2 \pi } \left(
    \frac{\eta_\chi m_b m_w \Delta m }{2 (4 m_w^2-m_H^2)} \right)^2
\right] \, \frac{1}{(4 m_w^2-m_h^2)^2 + m_h^2 \Gamma_h^2} \ \frac{3}{2
  x_{\rm f.o.}} \sim 2.5 \times 10^{-9} {\rm GeV}^{-2} \, .
\label{eq:RelD}
\eeq

To determine the allowed region of the parameter space, for $m_w =
50~{\rm GeV}$, we solve (\ref{eq:FermiBub}) and (\ref{eq:RelD}) while
simultaneously demanding that $\tilde{\alpha} \alt 0.01$, and that the
upper limit on the invisible decay width for the SM Higgs
(\ref{LHCwidth}) is not violated by (\ref{eq:invis}). The best--fit
parameters are given in Table~\ref{table:1}, for an example with
$\Delta m = 6000~{\rm GeV}$.  We can see that the
annihilation into pairs of Goldstone bosons is dominating the $w \bar
w$ interactions at freeze-out by a factor of about $9$. Precise
determination of the parameters is at present hampered by the large
uncertainties in the dark matter halo profile. Interestingly, the
$W$-WIMP-nucleon cross section is within the reach of the XENON1T
experiment~\cite{Aprile:2012zx}, providing a strong motivation for the ideas
discussed in this section.

\begin{table}
\caption{Best--fit parameters. \label{table:1}}
\begin{tabular}{ c  c }
\hline
\hline
$\Delta m$ & $6000~{\rm GeV}$ \\
\hline
$m_h$ & $98.8\ {\rm GeV}$ \\
\hline
$f$ & $0.34$ \\
\hline
$\tilde{\alpha}$ & $0.009$ \\
\hline
$\eta_\chi$ & $1.8 \times 10^{-4}$ \\
\hline
$\chi$ & $0.049$  \\
\hline
$\Gamma_{H \rightarrow \, {\rm invisible}}$ & $0.65~{\rm MeV}$ \\
\hline
$\quad \la \sigma_{\alpha'} v (x_{\rm f.o.}) \ra \quad$ & $\quad 2.7 \times 10^{-26} \ {\rm cm^3 s^{-1}} \quad $ \\
\hline
~~~~~~~~~~~~~~~$\la \sigma_b v(x_{\rm f.o.}) \ra$ ~~~~~~~~~~~~~~& ~~~~~~~~~~~~~~$0.3 \times 10^{-26} \ {\rm cm^3 s^{-1}} $~~~~~~~~~~~~~~\\
\hline
$\quad \la \sigma_{\alpha'}  v (x_{\rm G.h.}) \ra \quad$ & $7.8 \times 10^{-26} \ {\rm cm^3 s^{-1}}$ \\
\hline
\hline
\end{tabular}
\end{table}

\begin{figure}[tbp]
\includegraphics[width=0.9\textwidth]{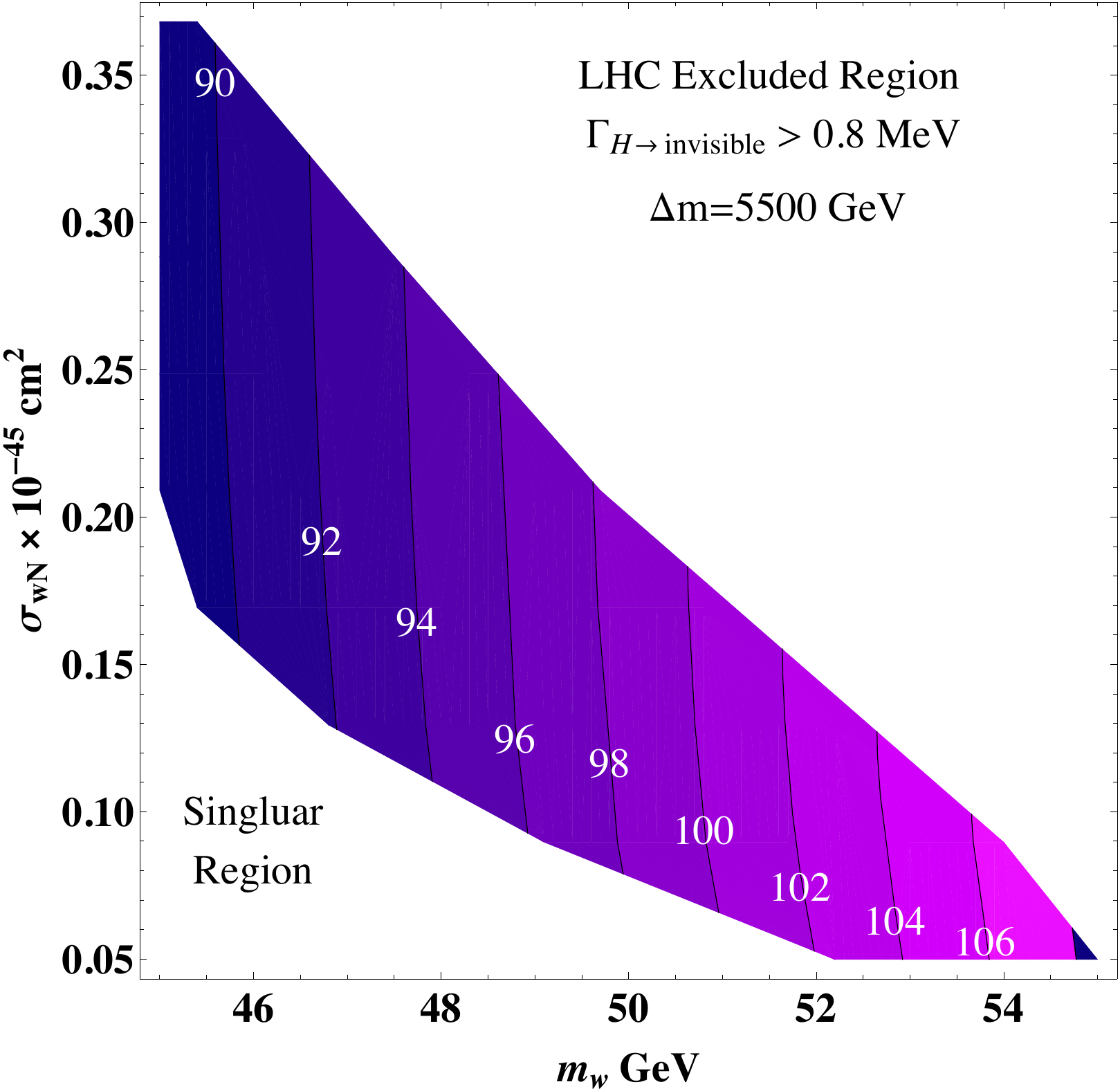}
\caption{ Contours of constant $m_h/{\rm GeV}$ in the $\sigma_{wN} - m_w$
  plane. The contours satisfy \emph{Fermi} data, the relic density
  requirement, and the LHC bound ${\cal B} (H \to \ {\rm invisible})$.
  We have required $\tilde{\alpha}\alt 0.01$ and taken $\Delta m = 5500~{\rm GeV}$.}
\label{fig:tres}
\end{figure}

\begin{figure}[tbp]
\includegraphics[width=0.9\textwidth]{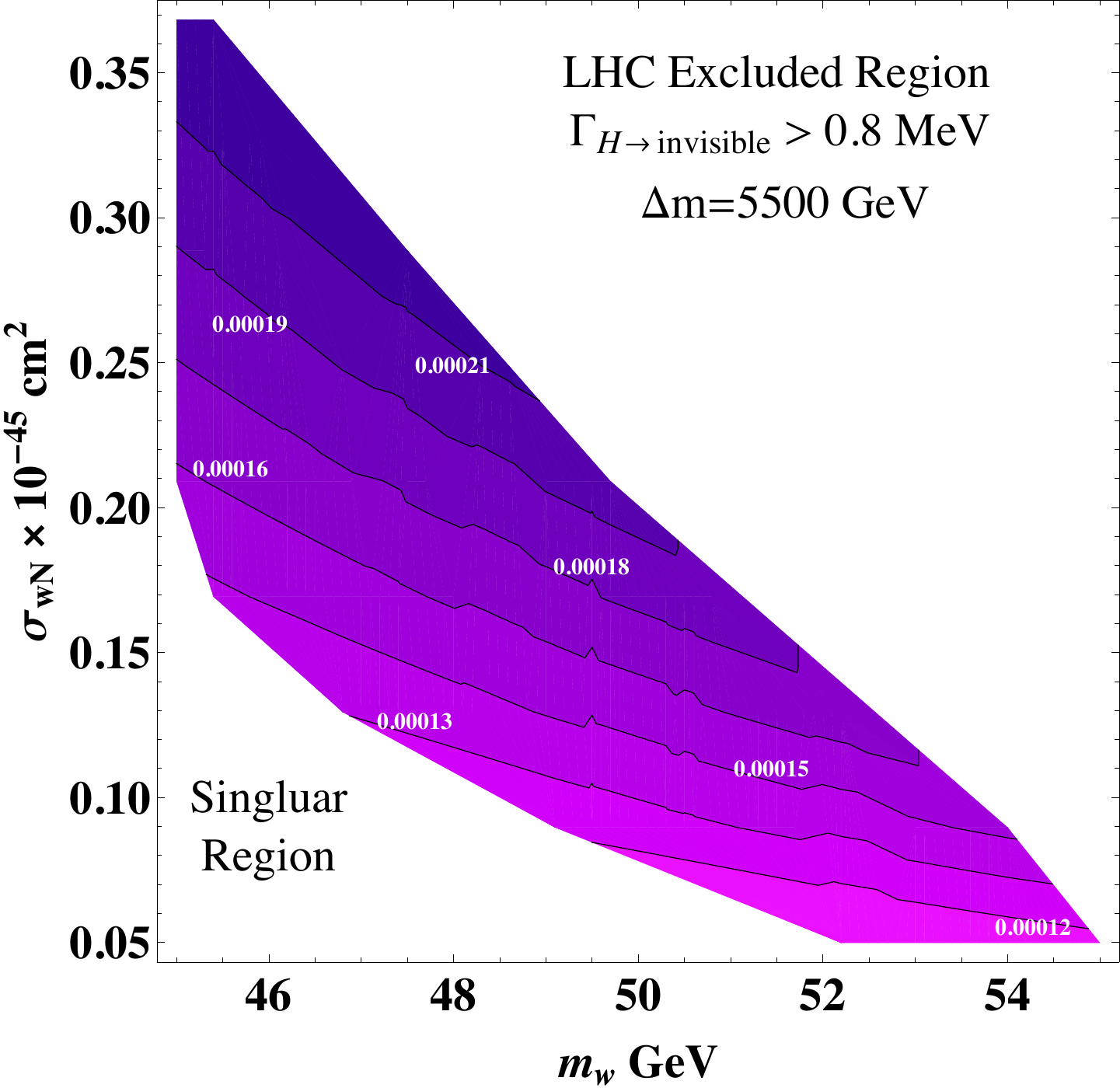}
\caption{Contours of constant $\eta_\chi$ in the $\sigma_{wN} - m_w$
  plane. Again the contours satisfy \emph{Fermi} data, the relic
  density requirement, and the LHC bound ${\cal B} (H \to \ {\rm
    invisible})$. We have required $\tilde{\alpha}\alt 0.01$ and taken
  $\Delta m = 5500~{\rm GeV}$.}
\label{fig:cuatro}
\end{figure}

\begin{figure}[tbp]
\includegraphics[width=0.9\textwidth]{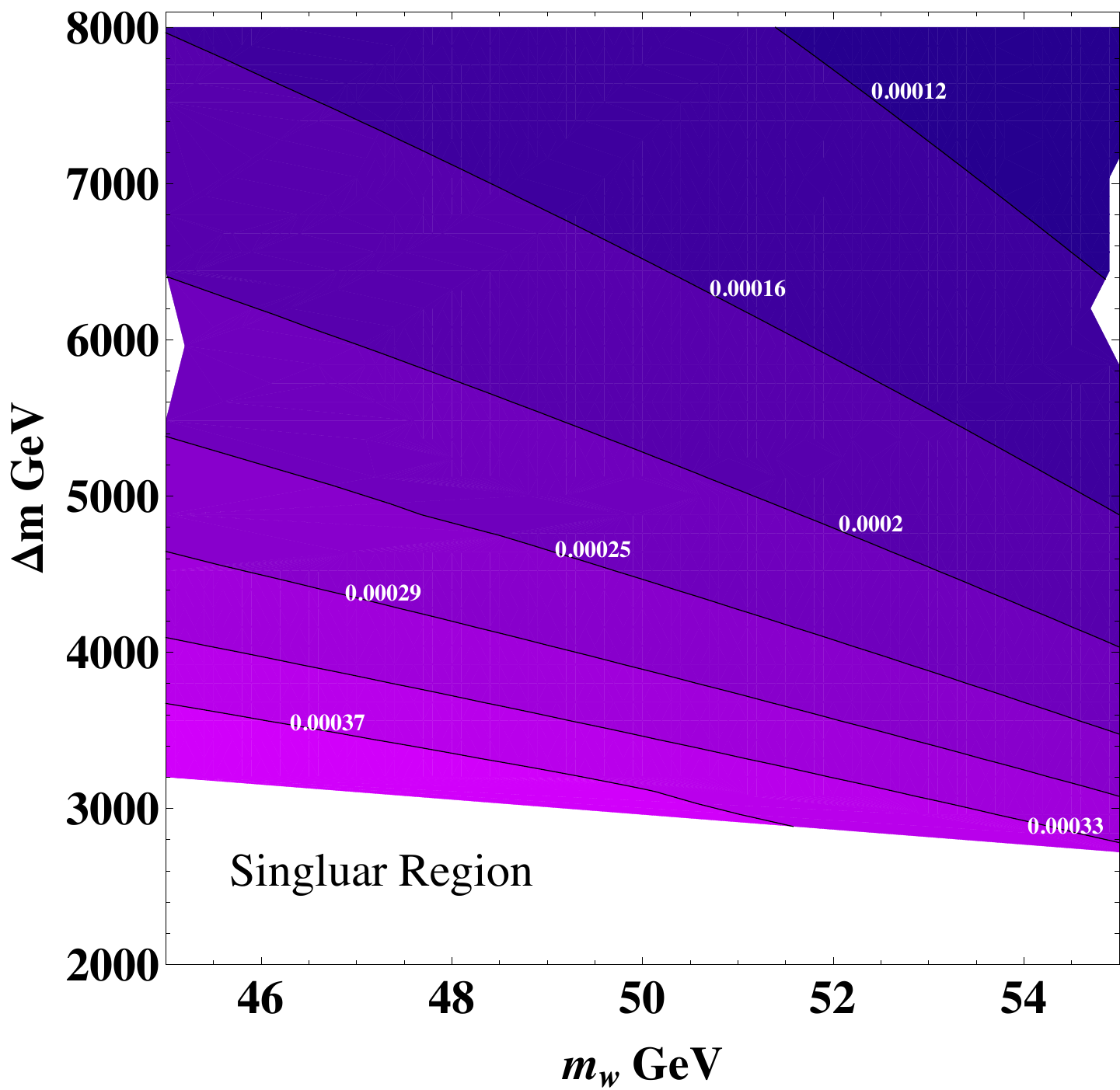}
\caption{ Contours of constant $\eta_\chi$ in the $\Delta m- m_w$
  plane. The contours satisfy \emph{Fermi} data, the relic density
  requirement, and saturates the LHC bound ${\cal B} (H \to \ {\rm
    invisible})$. We have required $\tilde{\alpha} \alt 0.01$ and we have verified
that the XENON-100 upper limit~\cite{Aprile:2011hi} is not violated.}
\label{fig:cinco}
\end{figure}

Duplicating the procedure described above, we have scanned the mass
range of the parameter space that is consistent with Fermi data:
$45~{\rm GeV} < m_w < 55~{\rm GeV}$; see Fig.~\ref{fig:dos}. Our
results are encapsulated in Figs.~\ref{fig:tres}, \ref{fig:cuatro} and
\ref{fig:cinco}.  In particular, Fig.~\ref{fig:tres} and
\ref{fig:cuatro} display, for $\Delta m = 5500~{\rm GeV}$, the region
of the parameter space of $m_w$ vs $\sigma_{wN}$ not yet
excluded by current direct detection experiments or the LHC.  Future
LHC data will either more tightly constrain this parameter space or
will turn up evidence for a signal.  Note that  the region
excluded by nonexistence of a solution ($\Gamma_{H \to \, {\rm
    invisible}} \approx 0.3~{\rm MeV}$) up to the current LHC bound
will be very tightly constrained after the LHC coming upgrade,
assuming no signal appears. In the case that a signal does appear, the
combination of relations shown in Figs.~\ref{fig:tres} and
\ref{fig:cuatro} will constrain model parameters providing the XENON1T
experiment with the specific cross section required to confirm this
model. As an illustration, in Fig.~\ref{fig:cinco} we show contours of
constant $\eta_\chi$ in the $\Delta m - m_w$ plane for the case in
which ${\cal B} (H \to\, {\rm invisible})$ saturates the current
limit, $\Gamma_{H \to \, {\rm invisible}} = 0.8~{\rm MeV}$. The direct
  detection cross section sampling this subregion of the parameter
  space varies between $1.8 \times 10^{-46}~{\rm cm}^2$ and $2.2
  \times 10^{-46}~{\rm cm}^2$, with an average of $1.9 \times
  10^{-46}~{\rm cm}^2$.

\section{$\bm{W}$-WIMP Interpretation for Hints of Light
  Dark Matter}

Signals broadly compatible with $\sim 10~{\rm GeV}$ dark matter have
been observed in four direct detection experiments:
DAMA/LIBRA~\cite{Bernabei:2010mq},
CoGeNT~\cite{Aalseth:2010vx,Aalseth:2011wp},
CRESST~\cite{Angloher:2011uu}, and CDMS-II~\cite{Agnese:2013rvf}. In
this section we explore the compatibility with one particular region
of the $W$-WIMP parameter space.  The features of this region of the
parameter space has bearing on the evidence for extrarelativistic
degrees of freedom at the CMB epoch.

In order to elaborate on the case for $m_w \sim 10~{\rm GeV}$, we
consider $m_h \approx 500~{\rm MeV}$ and $\eta_\chi \approx
0.005$. Substituting these values in (\ref{eq:DetAcdms}), it is
straightforward to see that to comply with the elastic cross section
signal reported by the CDMS Collaboration~\cite{Agnese:2013rvf}, we
must set $\Delta m \approx 1.75~{\rm GeV}$. This in turn determines
via (\ref{eq:TEWII}) a thermal average annihilation cross section into
quarks: $\langle \sigma_b v (x_{\rm G.h.}) \rangle \approx 1.3 \times
10^{-39}~{\rm cm}^3 \, {\rm s}^{-1}$. Note that this is more than 10
orders of magnitude smaller than current limits on light dark matter
from antiproton data~\cite{Lavalle:2010yw,Evoli:2011id,Kappl:2011jw}.

The observed dark matter density is obtained again through dominant $W$-WIMP
annihilation into the hidden sector.  To demonstrate this point, we
must first compute the $w \bar w  \rightarrow 2 h$ annihilation cross
section, as this channel is now open. We consider the relevant terms of (\ref{eq:fifty}), \beq
\frac{f \cos \chi}{2} h \bar \psi_- \psi_- + \frac{f \sin \chi}{2} H
\bar \psi_- \psi_- \ , \eeq as well as the relevant terms of the
scalar potential \beq \CV \approx \dots - \frac{m_h^2}{2 v_r} h^3
-\frac{\eta_\chi v_\phi}{2} \lpa \frac{m_H^2 + 2 m_h^2}{m_H^2 - m_h^2}
\rpa H h^2 \ ; \eeq together this gives the total reaction matrix
element
\begin{eqnarray} \CM  & = & i f
\bar{v}(p) u(p') \frac{i}{s-m_h^2} \lpa \frac{-i 3! m_h^2}{2v_r}\rpa +
i f \frac{\eta_\chi v_r v_\phi}{m_H^2-m_h^2} \bar{v}(p)
u(p')\frac{i}{s-m_H^2}  \nonumber \\
& \times &  \lpa \frac{-i \eta_\chi v_\phi
  (m_H^2+2m_h^2)}{m_H^2-m_h^2} \rpa \ . 
\end{eqnarray}
Assuming $\eta_\chi \ll 1$ and $m_h \ll m_H$, we arrive at a
manageable form of the spin-averaged $w \bar w \to 2h$ amplitude
\begin{eqnarray}
\frac{1}{4} \sum_{\rm spins}|\CM|^2 & \approx &  f^2 \lpa \frac{9
  m_h^4}{v_r^2 (s-m_h^2)^2} + \frac{6 m_h^2 \eta_\chi^2 v_\phi^2
  (m_H^2+2m_h^2)}{m_H^4 (s-m_h^2)(s-m_H^2)}\rpa (p\cdot p' - m_w^2) \ , \nonumber
\\
& \approx &  f^2 \lpa  \frac{18 f^2 m_h^4}{\Delta m^2 (s-m_h^2)^2} +
\frac{3 m_h^2 \eta_\chi^2 v_\phi^2}{m_H^2 (s-m_h^2)(s-m_H^2)}\rpa (s - 4m_w^2) \ , 
\end{eqnarray}
and the scattering cross section
\begin{equation}
\sigma \approx \frac{f^2}{32 \pi s }\sqrt{\frac{s-4m_h^2}{s-4m_w^2}}
\lpa  \frac{18 f^2 m_h^4}{\Delta m^2 (s-m_h^2)^2} + \frac{3 m_h^2
  \eta_\chi^2 v_\phi^2}{m_H^2 (s-m_h^2)(s-m_H^2)}\rpa (s - 4m_w^2) \ . 
\end{equation}
We take the thermal average in the low temperature limit, that is $T \ll m_w$,
\beq \la \sigma_h v \ra \approx \lpa \frac{9 f^4 m_h^4}{8 \pi \Delta
  m^2 (m_h^2-4m_w^2)^2}+\frac{3 f^2 v_\phi^2 \eta_\chi^2 m_h^2}{16 \pi
  m_H^4 (m_h^2-4 m_w^2)} \rpa \la v^2 \ra.
\eeq

\begin{figure}[tbp]
\includegraphics[width=0.9\textwidth]{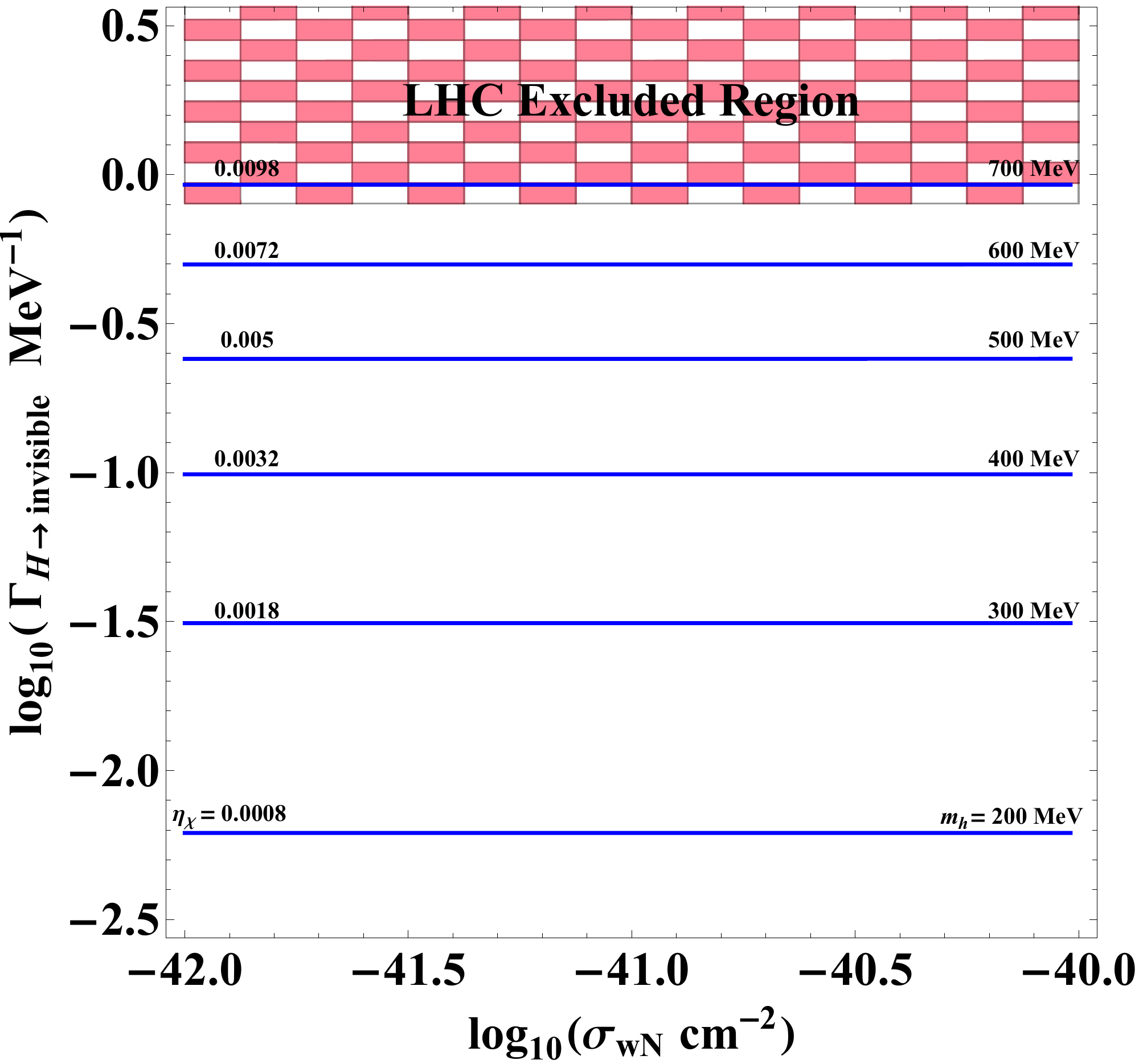}
\caption{$\Gamma_{H \rightarrow {\rm invisible}}$ for varying values
  of $\sigma_{wN}$. The plotted values are nearly constant as the
  terms from $\Gamma_{H \rightarrow \alpha, h}$ dominate the decay
  width, and thus there is weak dependence on the direct detection
  cross section. For varying values of $m_h$ we adjust the value of
  $\eta_\chi$ so that the Goldstone bosons decouple from the
  primordial plasma at $k_BT \approx m_\mu$, yielding $N_{\rm eff} =
  3.39$. For $200~{\rm MeV} \leq m_h \leq 700~{\rm MeV}$, the Higgs
  decay width into the hidden sector varies between $(0.006 -
  0.92)$~MeV. The constant $\eta_\chi/m_h$ contours shown here are
  independent of $m_w$, and therefore span the mass range $7~{\rm
    GeV} \alt m_w/{\rm GeV} \alt 10$.}
\label{fig:seis}
\end{figure}

By demanding the total annihilation cross section to comply with the
relic density requirement~\cite{Steigman:2012nb} we obtain \beq \la
\sigma_{\alpha'} v \ra_ + \la \sigma_h v \ra + \sum_{\rm fermions} \la
\sigma_i v \ra \sim 3 \times 10^{-26}~{\rm cm}^3/s \,,  \eeq and so
\begin{equation}
f \approx 0.04 \ .
\label{eq:WeinLaster}
\end{equation} 
As a final check we ensure that the LHC
upper limit on the hidden decay width of the Higgs is satisfied;
taking note that the decay channel $H \rightarrow \bar \psi_+ \psi_+$
is now open, we have 
\beq \frac{\eta_\chi^2 v_\phi^2}{16\pi m_H} +
\frac{\eta_\chi^2 \Delta m^2 v_\phi^2}{32\pi m_H^3}=0.24~{\rm MeV} <
0.8~{\rm \rm MeV} \, . \eeq In Fig.~\ref{fig:seis} we exhibit the
range of parameters consistent with the 95\%~C.L. upper limit on ${\cal
  B} (H \to {\rm
  invisible})$~\cite{Espinosa:2012vu,Cheung:2013kla,Giardino:2013bma,Ellis:2013lra}
together with possible signal regions associated with data from
CDMS-II~\cite{Agnese:2013rvf}. For $m_w = 10~{\rm GeV}$, the best-fit
intervals at the 68\%~C.L. and the 90\%~C.L. are $3 \times 10^{-42} <
\sigma_{wN}/{\rm cm}^2 < 2.5 \times 10^{-41}$ and $2 \times 10^{-42} <
\sigma_{wN}/{\rm cm}^2 < 3 \times 10^{-41}$, respectively. The
horizontal lines preserve the constant $\eta_\chi/m_h$ ratio that
allows decoupling of $\alpha'$ at $T \approx m_\mu$, yielding $N_{\rm
  eff} = 3.39$. 

In summary, we have shown that $W$-WIMPs of about 10~GeV can
simultaneously explain the observed relic density and the possible
signals observed by direct detection experiments, while avoiding
limits from indirect detection experiments. In the near future, the
Large Underground Xenon (LUX) dark matter
experiment~\cite{Akerib:2012ys} will collect enough statistics to
probe the $\sim 10~{\rm GeV}$ dark matter hypothesis. Concurrent with
LUX observations will be precise measurements of the Higgs branching
fractions by the LHC ATLAS and CMS experiments (operating at $\sqrt{s}
= 14~{\rm TeV}$). This new arsenal of data, when combined with
observations the Phased IceCube Next Generation Upgrade
(PINGU)~\cite{Koskinen:2011zz}, will have the potential to single out
this distinctive Higgs portal light dark matter model.\footnote{Since
  the annihilation rate into SM particles is largely suppressed
  compared to annihilations into the hidden sector, this particular
  model predicts null results at PINGU.}

\section{Conclusions}

$\gamma$-ray data in the $1 - 100~{\rm GeV}$ range from \emph{Fermi}
show a new unexpected feature of the Milky Way: two huge spheroidal
structures called \emph{Fermi} bubbles, extending up to about 10~kpc
($50^\circ$) out of the GC on either side of the Galactic disk.
Intriguingly, the bubbles coincide spatially with the WMAP haze in
microwave and the thermal x-ray emission seen by ROSAT. There is a
general consensus in that far from the GC the $\gamma$-rays observed
by \emph{Fermi} originate from inverse Compton scattering of photons
from the interstellar radiation field by the same hard electron
population that produces the haze via synchrotron. Very recently,
a second component of $\gamma$-ray emission from the low-latitude
regions of the \emph{Fermi} bubbles has been identified. The spectral
shape of this new component is consistent with that expected from an
approximately $50~{\rm GeV}$ dark matter particle annihilating into $b
\bar b$, with a normalization corresponding to $\langle \sigma_b v
\rangle \sim 8 \times 10^{-27}~{\rm cm}^3/{\rm s}$, or else with
$10~{\rm GeV}$ dark matter particle annihilating dominantly into
$\tau^+ \tau^-$, with a normalization corresponding to $\langle
\sigma_\tau v \rangle \sim 2 \times 10^{-27}~{\rm cm}^3/{\rm s}$.

We have shown that $W$-WIMPs (with $m_w \approx 50~{\rm GeV}$) are
capable of accommodating the desired effective annihilation into $b
\bar b$. We have also demonstrated that the thermal cross section
required to account for the relic dark matter abundance can easily be
obtained if $w \bar w \to 2 \alpha$ is the dominant annihilation
channel.  However, given that the Goldstone bosons would decouple at
$~5~{\rm GeV}$ ({\it i.e.} in the very early Universe), the
contribution to the effective number of neutrinos for the described
parameter space is negligible and thus cannot explain the evidence for
dark radiation (assuming there exist no common systematic uncertainties
in the measurements of the Hubble parameter).  In the near future, the
upgraded LHC together with the new XENON1T experiment will further
whittle down the parameter space, or else make a discovery.

If $m_w \approx 10~{\rm GeV}$, Weinberg's hidden sector does not
provide a viable explanation of the \emph{Fermi} bubbles since the
$W$-WIMP annihilation would be dominated by $b \bar b$ rather than
$\tau^+ \tau^-$.  However, there remains an interesting region of the
parameter space which can account for the alleged signals recently
reported by direct detection experiments.  In this region, the
Goldstone bosons decouple from the primordial plasma near the 100~MeV
temperature, consistent with the two measurements of the effective
number of neutrinos reported by the Planck Collaboration. In this
region of the parameter space, $W$-WIMP annihilation into Goldstone
bosons is also sufficient for consistency of the observed dark matter
abundance. Furthermore, future LHC measurements will further constrain
this sector of the Higgs portal (or bettter, find a signal), while LUX
will close the deliberations on the alleged direct signals.

In closing, we note that the \emph{Fermi} bubble production mechanism
proposed in this paper is also applicable to more general ``hidden
valleys''~\cite{Strassler:2006im,Heikinheimo:2013xua}. The mixing
between a more elaborate hidden sector and the visible sector
modulates the rate of annihilation into SM fields and hidden Goldstone
bosons.

\section*{Acknowledgments}
We thank Dan Hooper, Tom Paul, Diego Torres, and Tom Weiler for some
valuable discussion. L.A.A. and B.J.V. are supported by the
U.S. National Science Foundation (NSF) under CAREER Grant PHY-1053663.

\end{document}